\documentclass[%
 reprint,
superscriptaddress,
 amsmath,amssymb,
 aps,
]{revtex4-2}

\usepackage{graphicx}
\usepackage{dcolumn}
\usepackage{bm}
\usepackage{hyperref}
\usepackage[mathlines]{lineno}
\usepackage{url}
\usepackage[caption=false]{subfig}
\usepackage{float}
\usepackage{cleveref}
\usepackage{siunitx}
\usepackage[caption=false]{subfig}
\usepackage{ragged2e}

\begin{document}

\preprint{APS/123-QED}

\title{Astrophysical Sensitivity Projections for the IceCube Upgrade}
\affiliation{III. Physikalisches Institut, RWTH Aachen University, D-52056 Aachen, Germany}
\affiliation{Department of Physics, University of Adelaide, Adelaide, 5005, Australia}
\affiliation{Dept. of Physics and Astronomy, University of Alaska Anchorage, 3211 Providence Dr., Anchorage, AK 99508, USA}
\affiliation{School of Physics and Center for Relativistic Astrophysics, Georgia Institute of Technology, Atlanta, GA 30332, USA}
\affiliation{Dept. of Physics, Southern University, Baton Rouge, LA 70813, USA}
\affiliation{Dept. of Physics, University of California, Berkeley, CA 94720, USA}
\affiliation{Lawrence Berkeley National Laboratory, Berkeley, CA 94720, USA}
\affiliation{Institut f{\"u}r Physik, Humboldt-Universit{\"a}t zu Berlin, D-12489 Berlin, Germany}
\affiliation{Fakult{\"a}t f{\"u}r Physik {\&} Astronomie, Ruhr-Universit{\"a}t Bochum, D-44780 Bochum, Germany}
\affiliation{Universit{\'e} Libre de Bruxelles, Science Faculty CP230, B-1050 Brussels, Belgium}
\affiliation{Vrije Universiteit Brussel (VUB), Dienst ELEM, B-1050 Brussels, Belgium}
\affiliation{Dept. of Physics, Simon Fraser University, Burnaby, BC V5A 1S6, Canada}
\affiliation{Department of Physics and Laboratory for Particle Physics and Cosmology, Harvard University, Cambridge, MA 02138, USA}
\affiliation{Dept. of Physics, Massachusetts Institute of Technology, Cambridge, MA 02139, USA}
\affiliation{Dept. of Physics and The International Center for Hadron Astrophysics, Chiba University, Chiba 263-8522, Japan}
\affiliation{Department of Physics, Loyola University Chicago, Chicago, IL 60660, USA}
\affiliation{Dept. of Physics and Astronomy, University of Canterbury, Private Bag 4800, Christchurch, New Zealand}
\affiliation{Dept. of Physics, University of Maryland, College Park, MD 20742, USA}
\affiliation{Dept. of Astronomy, Ohio State University, Columbus, OH 43210, USA}
\affiliation{Dept. of Physics and Center for Cosmology and Astro-Particle Physics, Ohio State University, Columbus, OH 43210, USA}
\affiliation{Niels Bohr Institute, University of Copenhagen, DK-2100 Copenhagen, Denmark}
\affiliation{Dept. of Physics, TU Dortmund University, D-44221 Dortmund, Germany}
\affiliation{Dept. of Physics and Astronomy, Michigan State University, East Lansing, MI 48824, USA}
\affiliation{Dept. of Physics, University of Alberta, Edmonton, Alberta, T6G 2E1, Canada}
\affiliation{Erlangen Centre for Astroparticle Physics, Friedrich-Alexander-Universit{\"a}t Erlangen-N{\"u}rnberg, D-91058 Erlangen, Germany}
\affiliation{Physik-department, Technische Universit{\"a}t M{\"u}nchen, D-85748 Garching, Germany}
\affiliation{D{\'e}partement de physique nucl{\'e}aire et corpusculaire, Universit{\'e} de Gen{\`e}ve, CH-1211 Gen{\`e}ve, Switzerland}
\affiliation{Dept. of Physics and Astronomy, University of Gent, B-9000 Gent, Belgium}
\affiliation{Dept. of Physics and Astronomy, University of California, Irvine, CA 92697, USA}
\affiliation{Karlsruhe Institute of Technology, Institute for Astroparticle Physics, D-76021 Karlsruhe, Germany}
\affiliation{Karlsruhe Institute of Technology, Institute of Experimental Particle Physics, D-76021 Karlsruhe, Germany}
\affiliation{Dept. of Physics, Engineering Physics, and Astronomy, Queen's University, Kingston, ON K7L 3N6, Canada}
\affiliation{Department of Physics {\&} Astronomy, University of Nevada, Las Vegas, NV 89154, USA}
\affiliation{Nevada Center for Astrophysics, University of Nevada, Las Vegas, NV 89154, USA}
\affiliation{Dept. of Physics and Astronomy, University of Kansas, Lawrence, KS 66045, USA}
\affiliation{UCLouvain, Centre for Cosmology, Particle Physics and Phenomenology, CP3, Chemin du Cyclotron 2, 1348 Louvain-la-Neuve, Belgium}
\affiliation{Department of Physics, Mercer University, Macon, GA 31207-0001, USA}
\affiliation{Dept. of Astronomy, University of Wisconsin{\textemdash}Madison, Madison, WI 53706, USA}
\affiliation{Dept. of Physics and Wisconsin IceCube Particle Astrophysics Center, University of Wisconsin{\textemdash}Madison, Madison, WI 53706, USA}
\affiliation{Institute of Physics, University of Mainz, Staudinger Weg 7, D-55099 Mainz, Germany}
\affiliation{Department of Physics, Marquette University, Milwaukee, WI 53201, USA}
\affiliation{Institut f{\"u}r Kernphysik, Universit{\"a}t M{\"u}nster, D-48149 M{\"u}nster, Germany}
\affiliation{Bartol Research Institute and Dept. of Physics and Astronomy, University of Delaware, Newark, DE 19716, USA}
\affiliation{Dept. of Physics, Yale University, New Haven, CT 06520, USA}
\affiliation{Columbia Astrophysics and Nevis Laboratories, Columbia University, New York, NY 10027, USA}
\affiliation{Dept. of Physics, University of Oxford, Parks Road, Oxford OX1 3PU, United Kingdom}
\affiliation{Dipartimento di Fisica e Astronomia Galileo Galilei, Universit{\`a} Degli Studi di Padova, I-35122 Padova PD, Italy}
\affiliation{Dept. of Physics, Drexel University, 3141 Chestnut Street, Philadelphia, PA 19104, USA}
\affiliation{Physics Department, South Dakota School of Mines and Technology, Rapid City, SD 57701, USA}
\affiliation{Dept. of Physics, University of Wisconsin, River Falls, WI 54022, USA}
\affiliation{Dept. of Physics and Astronomy, University of Rochester, Rochester, NY 14627, USA}
\affiliation{Department of Physics and Astronomy, University of Utah, Salt Lake City, UT 84112, USA}
\affiliation{Dept. of Physics, Chung-Ang University, Seoul 06974, Republic of Korea}
\affiliation{Oskar Klein Centre and Dept. of Physics, Stockholm University, SE-10691 Stockholm, Sweden}
\affiliation{Dept. of Physics and Astronomy, Stony Brook University, Stony Brook, NY 11794-3800, USA}
\affiliation{Dept. of Physics, Sungkyunkwan University, Suwon 16419, Republic of Korea}
\affiliation{Institute of Physics, Academia Sinica, Taipei, 11529, Taiwan}
\affiliation{Dept. of Physics and Astronomy, University of Alabama, Tuscaloosa, AL 35487, USA}
\affiliation{Dept. of Astronomy and Astrophysics, Pennsylvania State University, University Park, PA 16802, USA}
\affiliation{Dept. of Physics, Pennsylvania State University, University Park, PA 16802, USA}
\affiliation{Dept. of Physics and Astronomy, Uppsala University, Box 516, SE-75120 Uppsala, Sweden}
\affiliation{Dept. of Physics, University of Wuppertal, D-42119 Wuppertal, Germany}
\affiliation{Deutsches Elektronen-Synchrotron DESY, Platanenallee 6, D-15738 Zeuthen, Germany}

\author{R. Abbasi}
\affiliation{Department of Physics, Loyola University Chicago, Chicago, IL 60660, USA}
\author{M. Ackermann}
\affiliation{Deutsches Elektronen-Synchrotron DESY, Platanenallee 6, D-15738 Zeuthen, Germany}
\author{J. Adams}
\affiliation{Dept. of Physics and Astronomy, University of Canterbury, Private Bag 4800, Christchurch, New Zealand}
\author{J. A. Aguilar}
\affiliation{Universit{\'e} Libre de Bruxelles, Science Faculty CP230, B-1050 Brussels, Belgium}
\author{M. Ahlers}
\affiliation{Niels Bohr Institute, University of Copenhagen, DK-2100 Copenhagen, Denmark}
\author{J.M. Alameddine}
\affiliation{Dept. of Physics, TU Dortmund University, D-44221 Dortmund, Germany}
\author{S. Ali}
\affiliation{Dept. of Physics and Astronomy, University of Kansas, Lawrence, KS 66045, USA}
\author{N. M. Amin}
\affiliation{Bartol Research Institute and Dept. of Physics and Astronomy, University of Delaware, Newark, DE 19716, USA}
\author{K. Andeen}
\affiliation{Department of Physics, Marquette University, Milwaukee, WI 53201, USA}
\author{C. Arg{\"u}elles}
\affiliation{Department of Physics and Laboratory for Particle Physics and Cosmology, Harvard University, Cambridge, MA 02138, USA}
% ...
\date{\today}

\collaboration{IceCube Collaboration}
\noaffiliation

\begin{abstract}
Embedded in the South Pole's glacial ice, IceCube detects neutrino-induced Cherenkov light using an array of digital optical modules equipped with single photomultiplier tubes (PMTs). The new extension installed in 2025/2026, the IceCube Upgrade, introduces densely instrumented multi‑PMT optical modules within the existing infill array known as IceCube DeepCore. It is expected to enhance sensitivity in the GeV regime, with commissioning of the detector expected to be complete by the end of 2026. We present the projected sensitivities of the IceCube Upgrade for three key analyses: neutrino transient searches, steady emission from point sources such as NGC~1068, and diffuse emission from the Milky Way. These case studies represent direct extensions of current IceCube analyses. Using new Monte Carlo datasets, we demonstrate that the IceCube Upgrade achieves order-of-magnitude improvement in sensitivity at low energies ($\lesssim\nobreak10$\,GeV) for time‑dependent sources across short timescales. Conversely, for time-independent searches, the relative impact of the IceCube Upgrade's low-energy data is diluted by the decade-long accumulation of high-energy archival data. 
Nevertheless, we project significant improvements for soft-spectrum sources especially across the southern sky, driven by the IceCube Upgrade's superior background rejection capabilities.
The improved sensitivity at low energies for both transient and steady sources will open up an expanded discovery window for IceCube in the GeV band over the next decade.
\end{abstract}

\maketitle

\section{\label{sec:Intro}Introduction}

Neutrinos interact only weakly with matter, making them unique cosmic messengers as they travel unperturbed from their astrophysical sources to Earth, even from dense environments that are opaque to electromagnetic radiation.
This property makes them an excellent probe of high-energy processes in astrophysical sources in the era of multi-messenger astronomy.
The detection of the diffuse neutrino flux by IceCube demonstrated that the Universe can be probed with high-energy neutrinos~\cite{IceCube:Diffuse2013}.
A variety of astrophysical sources, including both transient and steady emitters, have been proposed as possible contributors to this flux~\cite{annurev_Meszaros2017}.

Transient sources such as gamma-ray bursts (GRBs) and supernovae (SNe) have long been considered promising candidates for astrophysical neutrino emission~\cite{GRBNu1997PRL, CCSNNu_Murase2011PRD}. Searches for neutrino emission from these sources have constrained their contribution to the observed TeV--PeV neutrino flux~\cite{IceCube:GRB2022, IceCube:SN_2023}.
While these high-energy searches are essential, many transients are also predicted to emit neutrinos at lower energies, down to the GeV scale~\cite{GeVTransients_Sherman2025}. Transient sources are particularly well suited for low-energy searches, as their short durations significantly reduce the impact of the dominant background of atmospheric neutrinos~\cite{TimdepPSSearch_BRAUN2010}.
Although no significant neutrino emission has been detected in past searches at sub-TeV energies~\cite{GRECO_nova2023, GRECO_GRB2024, GRECO_GW2024}, some models predict distinct signatures at GeV energies, thereby motivating the extension of transient searches to this energy range.

In certain GRB scenarios, quasi-thermal neutrino emission has been predicted to arise from inelastic collisions between protons and neutrons following their decoupling in the relativistic outflow~\cite{GRBQTNu2000Bahcall, GRBQTNu2000Meszaros}.
The typical energy of quasi-thermal neutrinos is estimated to be $E_\nu\approx30\left(\Gamma/100\right)\left(\Gamma_{\rm rel}/3\right)\rm\,GeV$, where $\Gamma$ is the bulk Lorentz factor of the GRB jets and $\Gamma_{\rm rel}$ is the relative Lorentz factor between the protons and neutrons~\cite{Murase2013}.
This makes low-luminosity GRBs (LLGRBs) with potentially low Lorentz factors ($\Gamma\sim\mathcal{O}$(10)) excellent targets for GeV neutrino observations.
Quasi-thermal neutrinos can probe subphotospheric processes in GRBs that are inaccessible to electromagnetic observations, providing unique insights into the emission mechanism as well as the jet composition and acceleration.
Recently, gamma-ray-dim GRBs have also been suggested as promising GeV--TeV neutrino sources~\cite{Nakama2025GRB}.

Core-collapse SNe are also compelling sources of neutrinos from GeV to PeV energies~\cite{InteractSN_Murase2024}. In particular, the production of quasi-thermal neutrinos in the GeV energy range is expected when a rapidly rotating, strongly magnetized proto-neutron star (PNS) is formed, as it can launch a relativistic outflow that leads to proton-neutron collisions~\cite{MuraseSN2014, Carpio2024}.
Detection of such neutrinos would provide a unique opportunity to study SN outflows and shed light on the GRB–SN connection~\cite{SN:ReviewNu2018}.

Another promising class of GeV neutrino sources is novae. About 20 novae have been detected in GeV gamma rays by Fermi-LAT~\cite{Nova:Review2020Chomiuk}, among which the 2021 outburst of RS Ophiuchi (RS~Oph) was the first to be detected in very-high-energy gamma rays~\cite{magic2022rsoph, HESS2022RSOph, LST2025RSOph}. These gamma-ray observations suggest cosmic-ray acceleration at shocks in novae, which naturally motivates searches for accompanying neutrinos at comparable energies. Detecting such neutrinos would further support the hadronic origin of the gamma-ray emission from novae.

In addition to transient phenomena, steady astrophysical sources provide compelling targets for GeV neutrino observations.
IceCube has recently reported evidence for high-energy neutrino emission from NGC~1068~\cite{IceCube:NGC1068} and the Galactic plane (GP)~\cite{GP_Science_Paper}.
While these sources have so far been observed at TeV to PeV energies, it is of interest to extend searches to lower energies.
This is especially strongly motivated for NGC 1068, where the most recent IceCube analysis~\cite{IceCube:NGC1068_FollupUp} fits a soft neutrino spectrum ($\gamma = 3.4$) extending down to sub-TeV energies. While the observed emission is largely interpreted as high-energy protons interacting with ambient radiation fields and gas in the vicinity of the supermassive black hole~\cite{Saurenhaus2026, Carpio_Seyfert_2026, Eichmann2026BayesianSeyfert}, the exact interaction channels and acceleration mechanisms remain uncertain~\cite{NGC1068Review}. Probing the currently unconstrained GeV band is essential for testing these competing scenarios.
The GP represents another well-motivated target for GeV neutrino observations. IceCube has observed TeV neutrino emission in this region~\cite{GP_Science_Paper}, which is expected to extend down to GeV energies, overlapping with the Galactic diffuse gamma-ray emission measured by Fermi-LAT~\cite{FermiLAT:GP}. GeV neutrino observations thus potentially allow for a direct comparison with the gamma-ray flux, probing the underlying cosmic-ray interactions.

These science cases highlight the importance of improved detector instrumentation with enhanced sensitivity to neutrinos at GeV energies.
This work presents sensitivity projections for the IceCube Upgrade and discusses its astrophysical potential. \Cref{sec:IceCube} introduces the detector, while \Cref{sec:datasets} details the simulation and reference datasets. \Cref{sec:Framework} outlines the analysis framework.
The results are presented and discussed in \Cref{sec:Results}, followed by a discussion of the impact of the deployed detector configuration in \Cref{sec:IC91}.
We conclude with an outlook in \Cref{sec:Conclusion}.

\section{\label{sec:IceCube}IceCube and the IceCube Upgrade}

The primary challenge in neutrino astronomy is the small interaction cross-section of neutrinos ($\mathcal{O}(10^{-34}\,\text{cm}^2)$ at 100\,TeV)~\cite{IceCube:CrossSections}, which necessitates large detection volumes.
The IceCube Neutrino Observatory utilizes 1\,km$^3$ of optically transparent, deep Antarctic glacial ice as its detection medium~\cite{IceCube:IceCube}.
Neutrinos interacting with an ice nucleus produce relativistic charged secondary particles that emit a cone of Cherenkov radiation~\cite{Theory:Cherenkov}, resulting in an indirect detection of neutrinos.
This optical signature is registered by a three-dimensional array of optical modules housing photomultiplier tubes (PMTs)~\cite{IceCube:PMTPaper}.

IceCube has been fully operational since 2011. Its optical array contains 5,160 Digital Optical Modules (DOMs)~\cite{IceCube:DOM} distributed across 86 vertical strings. Each DOM contains a single, downward-facing 10-inch PMT. The array's 125\,m horizontal and 17\,m vertical spacing is optimized for detecting neutrinos in the TeV--PeV energy range, with an energy threshold of $\sim 100$\,GeV.
Located in the deep, clear ice in the central lower part of the detector, IceCube DeepCore~\cite{IceCube:DeepCoreDesign2012} is a densely instrumented sub-detector comprising eight of the 86 strings, which feature a 7\,m vertical DOM spacing and an inter-string spacing of 41--105\,m. The denser instrumentation coupled with closer string-to-string spacing allows DeepCore to resolve neutrino events down to energies of $\sim10$\,GeV. Historically, this has enabled low-energy-focused analyses, such as measurements of neutrino oscillation parameters \cite{Abbasi2023,nu_osc_deepcore}.

\begin{figure}[htb]
    \centering
    \includegraphics[width=0.95\linewidth]{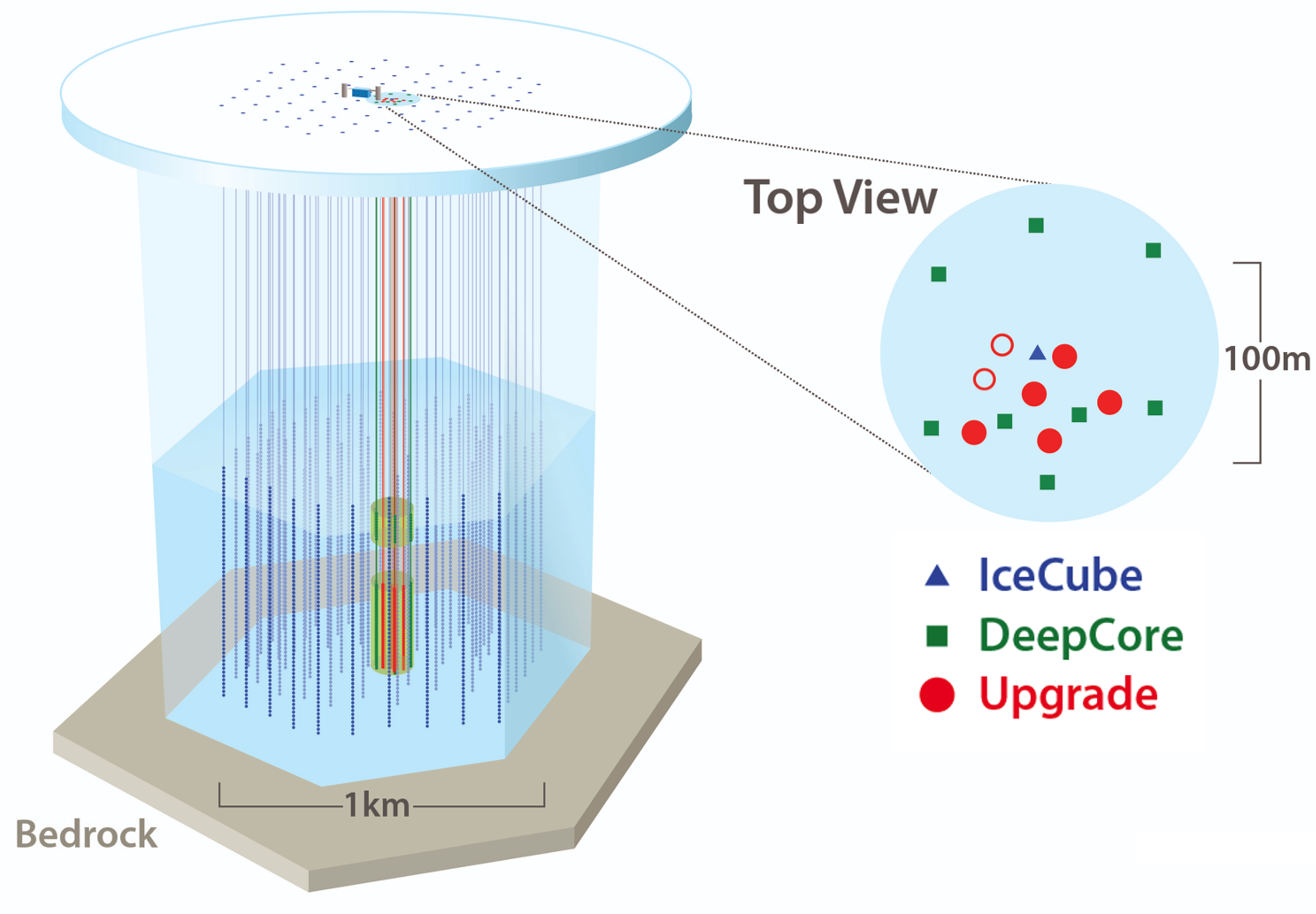}
    \caption{Detector overview of existing IceCube (blue) and DeepCore (green) as well as the originally planned IceCube Upgrade (red) geometry, with open circles indicating missing strings in the final configuration.}
    \label{fig:Upgrade_Overview}
\end{figure}

\begin{figure*}[tb]
    \centering
    \includegraphics[width=0.95\textwidth]{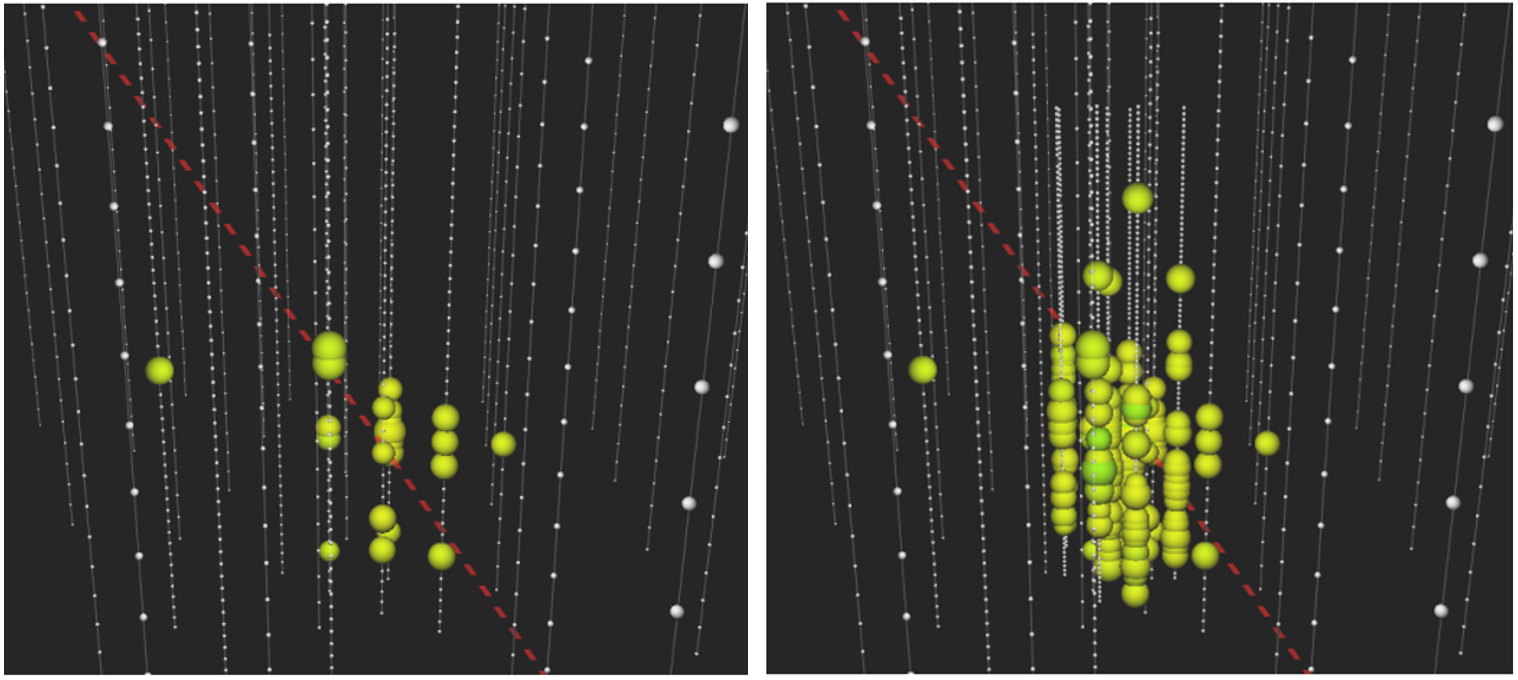}
    \caption{Event display of the same 30\,GeV $\nu_\mu$ CC interaction in DeepCore (left) and the IceCube Upgrade (right). Colored spheres correspond to module hits and the red line indicates the incident neutrino path. Taken from~\cite{Stuttard2020}.}
    \label{fig:Upgrade_EventDisplay}
\end{figure*}

During the 2025-2026 polar summer season, the IceCube Upgrade has been deployed, which represents the next low-energy extension of the IceCube detector. As shown in~\Cref{fig:Upgrade_Overview}, it was designed to instrument seven new strings with approximately 700 newly developed optical modules within the DeepCore volume, whereas the final deployment comprises five strings. Note that the dataset used in this paper reflects the initially planned seven-string geometry, distinct from the actual five-string deployment.
The instrumentation geometry features a vertical spacing of 3\,m and a horizontal inter-string spacing of $\sim$20--30\,m, representing the densest instrumented region in IceCube. This configuration is designed to fully contain and precisely reconstruct $\mathcal{O}$(GeV) neutrino interactions~\cite{IceCube:QuesoPaper}.
At GeV energies, neutrino–nucleus interactions are dominated by quasi-elastic scattering, resonance production, and deep inelastic scattering.
Charged-current (CC) interactions of $\nu_\mu$ create penetrating muons that leave elongated, track-like signatures in the detector. In contrast, CC interactions of other flavors and neutral-current interactions produce localized particle showers, resulting in a roughly spherical event morphology referred to as a cascade.

Apart from the denser instrumentation, the performance gain is driven by new optical modules: the mDOM (multi-PMT Digital Optical Module)~\cite{IceCube:mDOM}, with 24 3-inch PMTs, and the D-Egg~\cite{IceCube:DEgg}, which houses two 8-inch PMTs. This new hardware increases photon collection per optical module by a factor $>2.5$~\cite{IceCube:Gen2TDR} and remedies the limited, downward-focussed field of view of the original DOMs. mDOMs further provide a near-uniform 4$\pi$ angular acceptance and allow for intrinsic directional information at module level.

The IceCube Upgrade's capabilities are illustrated in~\Cref{fig:Upgrade_EventDisplay}, which contrasts a simulated 30\,GeV $\nu_{\mu}$ CC interaction as seen by DeepCore (left) with the event resolved by the IceCube Upgrade (right). The IceCube Upgrade event display shows a substantially increased photon hit density and a more clearly resolved morphology, enabling better reconstruction of such low-energy events.

In addition to enhancing capabilities at GeV energies, the IceCube Upgrade is also designed to provide a high-precision calibration of the entire detector.
This is achieved through standalone calibration devices installed on the IceCube Upgrade strings~\cite{IceCube:POCAM, IceCube:PencilBeam, IceCube:UpgradeCameraSystem}, as well as dedicated calibration components integrated into each optical module. The resulting improvement in calibration enables an unprecedented in-situ characterization of the ice properties, which currently constitute the dominant systematic uncertainty. This enhanced calibration can also be applied to archival data, leading to improved angular resolution in past high-energy analyses.

With its dense instrumentation, the IceCube Upgrade provides exceptional sensitivity to GeV neutrinos, with its potential for atmospheric neutrino oscillations demonstrated in~\cite{IceCube:QuesoPaper, IceCube:QuesoPaper_ICRC}. Here, we investigate the expected impact of the IceCube Upgrade on astrophysical neutrino searches in this work. The improved detection capabilities at low energies, in conjunction with superior background rejection, are expected to enhance overall sensitivity, which is quantified using three case studies that follow prominent IceCube analyses: transient searches~\cite{IceCube:GRB2022, GRECO_GRB2024}, the point source NGC 1068~\cite{IceCube:NGC1068, IceCube:NGC1068_FollupUp}, and the extended emission from the GP~\cite{GP_Science_Paper}.

\section{\label{sec:datasets}IceCube Upgrade Simulated Data and Reference Datasets}

A simulated dataset referred to as ``IC93'' was developed in previous studies exploring the IceCube Upgrade’s physics potential for atmospheric neutrino oscillations~\cite{IceCube:QuesoPaper}.
In this work, we employ the same IC93 sample to evaluate the projected sensitivity of the IceCube Upgrade to astrophysical neutrino sources.
The dataset reflects the initially planned seven-string geometry, rather than the actual five-string deployment.
A comprehensive description of the IC93 sample is provided in~\cite{IceCube:QuesoPaper}.
Here, we briefly summarize the key procedures used to construct the dataset.

\begin{figure*}[t]
    \centering
    \subfloat[\justifying Sky-averaged, all-flavor summed, and $\nu + \bar{\nu}$ averaged effective area. The IC93 sample is simulated in the energy range 1--500\,GeV. The IC93 acceptance is smaller than that of GRECO at energies above $\sim100$\,GeV due to different event cuts between the two datasets. The ELOWEN effective area is adopted from~\cite{IceCube:ELOWEN_O4}.\label{fig:datasets_effarea}]{%
        \includegraphics[width=0.48\textwidth]{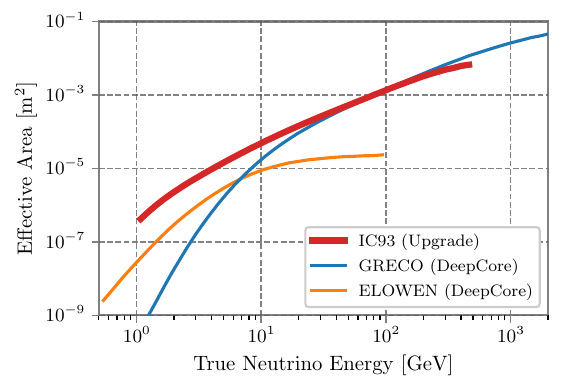}%
    }
    \hfill
    \subfloat[\justifying Angular resolution evaluated as the median and 68\,\% percentile of angular distance between true and reconstructed directions. The comparison is shown over the IC93 simulated energy range, while GRECO extends to energies up to $\sim10$\,TeV. The hatched regions indicate the kinematic limits evaluated from the retained events in each dataset.\label{fig:datasets_angres}]{%
        \includegraphics[width=0.48\textwidth]{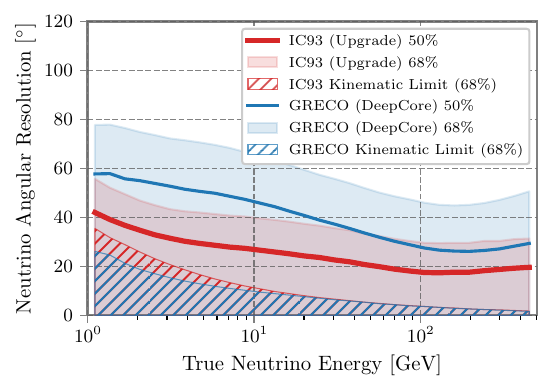}%
    }
    \caption{Key performance metrics of the IC93 sample and the existing DeepCore datasets.}
    \label{fig:datasets_properties}
\end{figure*}

Monte Carlo (MC) simulations for the IceCube Upgrade are generated following a procedure similar to that used for the DeepCore simulation in previous studies, with the IceCube Upgrade strings additionally implemented within the DeepCore volume.
Neutrino interactions are simulated using GENIE version 2.12.8~\cite{ANDREOPOULOS2010, ANDREOPOULOS2015}, with interaction vertices distributed in a cylindrical volume encompassing the DeepCore detector.
The energy of the injected neutrinos ranges from 1\,GeV to 500\,GeV.
Atmospheric muon events are generated with MuonGun~\cite{IceCube:MuonGun}, a toolkit for efficient muon simulation based on~\cite{BECHERINI2006}.
The propagation of particles produced in the interactions is simulated using PROPOSAL~\cite{KOEHNE2013} for muons and GEANT4~\cite{AGOSTINELLI2003} for all other particle types.
The Cherenkov photons produced by those charged particles are then propagated through the ice using CLSim~\cite{clsim}.

Lastly, the detector response is simulated to convert the Cherenkov photons reaching the optical modules into a series of readout pulses.
The pulses are individual signal outputs from the PMTs, characterized by their charge and time.
Noise pulses, consisting of the intrinsic PMT dark rate and scintillation and Cherenkov light from radioactive decays in the optical module's pressure vessel, are also added to the signal pulses.
Pulses that are correlated in both space and time are tagged as \textit{local coincident} (LC).
A trigger is issued if eight or more LC pulses are observed in the IceCube Upgrade within a 1.75\,\textmu s time window, thereby defining an individual event.

The triggered events are processed through the IC93 filter, adapted from the DeepCore filter~\cite{IceCube:DeepCoreDesign2012}.
The filter first cleans the pulses by discarding those that are isolated in space and time.
Next, a muon-veto algorithm is applied to the cleaned pulses to reject muon events while retaining neutrino events.
The filtering is followed by pulse cleaning using Graph Convolutional Neural Networks (GNNs)~\cite{GNNModel} and a series of simple, one-dimensional cuts.
These cuts are applied to basic variables, such as the number of modules with pulses and the number of pulses in the fiducial volume, primarily to suppress pure noise events.
Finally, the events are passed through a series of machine-learning classifiers to further reduce the atmospheric muon background.

Event reconstruction for the IC93 sample, covering the neutrino energy and direction as well as the associated uncertainty in the reconstructed direction, is performed using GNNs.
The GNN models take pulse information as input and reconstruct event parameters~\cite{IceCube:QuesoPaper}.
The per-event angular uncertainty $\sigma$ is derived from the concentration parameter $\kappa$ of the three-dimensional von Mises-Fisher distribution on the unit sphere, as $\sigma = 1/\sqrt{\kappa}$.
The models are trained and applied using GraphNet, an open-source deep learning library for neutrino telescopes~\cite{Sogaard2023}.

We construct background mock data for the IC93 dataset by randomly sampling neutrino events from the IC93 signal simulation, weighted according to the atmospheric flux~\cite{Honda2015AtmoNu}.
The resulting all-sky atmospheric neutrino background rate is 6.6\,mHz.
For each simulated neutrino event with an assigned weight $r$, which corresponds to its expected contribution to the total event rate, we sample the event following a Poisson distribution $P(n;rT)$, where $n$ indicates how many times the event is included in the mock data and $T$ is the assumed livetime.
The IC93 mock datasets comprise all neutrino flavors, including both neutrinos and antineutrinos, and are generated for different livetimes, ranging from 1 to 5 years.

To compare the IC93 sample with the current DeepCore datasets, we primarily use the GeV Reconstructed Events with Containment for Oscillation (GRECO) Astronomy dataset as the baseline.
The GRECO dataset, originally developed for neutrino oscillation studies with DeepCore~\cite{IceCube:Greco_intro}, has been modified into GRECO Astronomy through updated simulations and reoptimization of the event selection for sub-TeV transient searches~\cite{GRECO_nova2023}.
It has since been used in searches for various types of transients in this energy range~\cite{GRECO_GRB2024, GRECO_GW2024}.
The GRECO Astronomy dataset extracts sub-TeV neutrino candidates by applying a sequence of event selections, including a muon veto and simple cuts, followed by muon rejection and event reconstruction using boosted decision trees.
Per-event angular uncertainties associated with the reconstructed direction are derived using a random forest regressor.
GRECO Astronomy has an all-sky background rate of 4.6\,mHz, dominated by atmospheric neutrinos ($\mathord{\sim}60\,\%$) and muons ($\mathord{\sim}40\,\%$).
Details of GRECO Astronomy can be found in~\cite{GRECO_nova2023}.
The GRECO data used in this work spans April 26, 2012 to November 28, 2023, corresponding to a total livetime of 11.6~years.

Another DeepCore dataset, the Extremely Low-Energy (ELOWEN) sample~\cite{ELOWEN2021}, developed for transient searches with DeepCore at GeV energies, is also compared with the IC93 sample to evaluate the improvement in GeV transient sensitivity achieved with the IceCube Upgrade.
ELOWEN searches for GeV neutrinos by selecting extremely dim events in DeepCore that are still consistent with a physical origin. Rather than performing event reconstruction, ELOWEN searches for variations in the event rate of the selected sample, which has a typical rate of about 20\,mHz and is dominated by pure noise events.
Details are described in~\cite{ELOWEN2021}.

\renewcommand{\arraystretch}{1.2}
\begin{table}[tbp]
    \centering
    \caption{
    Expected all-sky background and signal event counts for the IceCube Upgrade and DeepCore datasets in a 1000\,s observation, assuming the GRB quasi-thermal neutrino emission model \cite{Murase_2022} with $\Gamma = 10$ and 30, scaled to a proxy distance of 1\,Mpc~\cite{Murase_2022}. The bolometric nucleon energy is assumed to be $\mathcal{E}_N = 5 \times 10^{50}$\,erg, corresponding to a neutrino fluence of $\sim 0.4$\,erg\,cm$^{-2}$. The IC93 background consists only of atmospheric neutrinos.}
    \label{tab:rates}
    \begin{tabular}{cccc}
    \hline
    Dataset & IC93 & GRECO & ELOWEN \\
    \hline
    Background Count & 6.6 & 4.6 & 20 \\
    Signal Count ($\Gamma=10$) & 1.8 & 0.10 & 0.29 \\
    Signal Count ($\Gamma=30$) & 5.4 & 1.4 & 0.95 \\
    \hline
    \end{tabular}
\end{table}

\Cref{fig:datasets_properties} compares key performance metrics for the IC93 (Upgrade), GRECO (DeepCore), and ELOWEN (DeepCore) datasets.
As shown in \Cref{fig:datasets_effarea}, the IC93 dataset significantly enhances the effective area at energies below $\sim 30$\,GeV, achieving a gain of about two orders of magnitude near 1\,GeV compared to GRECO.
Relative to ELOWEN, the IC93 sample improves the effective area by a factor of 5--10 in the GeV range.
As shown in \Cref{fig:datasets_angres}, the IC93 reconstruction, which includes IceCube Upgrade strings, consistently outperforms the GRECO dataset across the entire energy range, which is expected to further enhance the overall sensitivity for point-like sources. The ELOWEN dataset is not included in this comparison, as it does not involve any directional reconstruction.

The kinematic limits, defined as the 68\,\% percentile of the physical scattering angle between the incoming neutrino and the outgoing lepton, are indicated by the hatched regions.
The figure suggests that the angular resolution of the IC93 sample near 1\,GeV is non-negligibly constrained by the kinematic limit.
The differences in the apparent kinematic limits between the IC93 and GRECO datasets below 10 GeV arise from the opening-angle dependence of the signal efficiency, which results in different opening-angle distributions of the selected events in the two datasets.

Translating these instrumental improvements into expected event rates, \Cref{tab:rates} summarizes the expected numbers of all-sky background events and signal events for a 1000\,s observation in each dataset, adopting the quasi-thermal neutrino emission model for GRBs as a signal proxy~\cite{Murase_2022}.
The model features a peaked spectral shape around $E_\nu\approx0.2 \, \Gamma$\,GeV, as can also be seen in \Cref{fig:transient_diff_sens}.
The table shows that the IC93 sample provides a significantly enhanced signal acceptance at GeV energies compared to the DeepCore datasets, while the background count remains comparable to that of GRECO.
These features are expected to lead to an improvement in sensitivity, as quantified in the following sections.
Note that, for IC93 and GRECO, the effective background counts are further reduced by exploiting spatial information when observing point-like sources with known positions.
We only consider atmospheric neutrinos for IC93 here, as they constitute the dominant background contribution, as described below.

It is important to point out that the IC93 dataset, as used in this work, has a few key limitations.
The simulation is restricted to a true neutrino energy range of 1\,GeV to 500\,GeV, as seen in~\Cref{fig:datasets_effarea}, whereas the GRECO dataset extends to $\sim10$\,TeV. Consequently, by neglecting contributions from higher energies, the projected sensitivity of the IC93 sample presented here is conservative.
The IC93 sample also has limited statistics, which can create high-weight events in mock data. Therefore, the projections in this work are limited to a 5-year livetime.
Finally, the IC93 mock background data consists solely of atmospheric neutrinos and does not include contributions from muons or pure noise events.
This is because the statistics of muons and noise events in the current IC93 sample are insufficient to construct reliable background probability density functions (PDFs).
However, including muons and noise events is not expected to significantly affect the results presented here.
The estimated event rates of muons and pure noise events in the IC93 sample are approximately 0.8\,mHz and 0.2\,mHz (one event in the simulation with a livetime of 5000\,s), respectively, whereas the atmospheric neutrino rate is expected to be 6.6\,mHz.
Therefore, the contribution of muons and pure noise events to the total background rate is estimated to be approximately 15\,\%.
This is expected to reduce the sensitivity by only a factor of $\sqrt{1.15}\approx1.07$, in the background-dominated time regime.
We emphasize, however, that the background rates must ultimately be validated using real data from the IceCube Upgrade.

\section{\label{sec:Framework}Analysis Framework}
In this work, we quantify our sensitivity using the maximum-likelihood method, which is a standard approach used in the search of point-like or extended neutrino sources~\cite{PS_analysis_methods}. Since detector data is dominated by background events from atmospheric neutrinos and atmospheric muons, the aim of such analyses is to find an excess of signal-like neutrino events over the expected background.

\subsection{\label{sec:Framework_likelihoods}Unbinned Maximum Likelihood}

Using the unbinned maximum-likelihood method, the null (background-only) hypothesis is compared to an alternative (background-plus-signal) hypothesis.
In the latter case, the signal is assumed to follow a power-law spectrum $\Phi \propto E^{-\gamma}$ for the emission from a source.
The full likelihood function, which takes into account the Poisson fluctuations in the total number of events, is given by
\begin{align}
    L(n_s, \gamma) &= \frac{(n_s + n_b)^N e^{-(n_s+n_b)}}{N!} \notag \\
    &\quad \times \prod_{i=1}^{N} \left[\frac{n_s}{n_s+n_b} \mathcal{S}(x_i; \gamma) + \frac{n_b}{n_s + n_b} \mathcal{B}(x_i) \right], \label{eq:likelihood}
\end{align}
where $n_s$ is the number of signal events, $n_b$ is the expected number of background events, $N$ is the total number of events, and $\gamma$ is the spectral index of the signal events. Both $n_s$ and $\gamma$ are free parameters to be fitted.
The functions $\mathcal{S}$ and $\mathcal{B}$ denote the signal and background PDFs, respectively, which are used to evaluate the contribution of each event given its observables $x_i$. The observables include the reconstructed declination $\delta_i$, right ascension $\alpha_i$, angular uncertainty $\sigma_i$, and energy $E_i$ of the event.
The signal PDF $\mathcal{S}$ is factorized into a spatial PDF and an energy PDF.
The spatial PDF is modeled by a Kent distribution~\cite{Abbasi2024}, which describes the angular separation between the reconstructed event direction and the source position, taking the per-event angular uncertainty $\sigma_i$ as its spread.
The energy PDF represents the probability of an event having reconstructed energy $E_i$, given the assumed signal spectral index $\gamma$. 
The background PDF $\mathcal{B}$ depends only on the declination $\delta_i$ and the energy $E_i$ of the event, and is derived from data by scrambling the right ascension of the events~\cite{PS_analysis_methods}.

\Cref{eq:likelihood} is used for transient analyses, where the fluctuations in $N$ are relevant due to the limited event statistics.
Analyses of steady-state sources, on the other hand, use a large event sample, and therefore neglect the Poisson term and impose $n_s + n_b = N$.
For analyses using multiple datasets, as done for the time-independent analyses in this work, the likelihood functions from the individual datasets are multiplied to obtain an overall likelihood~\cite{PS_analysis_multiple_datasets}.
For the investigation of the GP, the signal-subtraction method is used to account for the contribution of signal events in the scrambled background data~\cite{Signal_subtraction}. This results in a slightly modified likelihood function, as described in Appendix~\ref{app:sig_sub}.

The test statistic (TS) is defined as
\begin{align*}
    \text{TS} = -2 \ln \left[\frac{L(n_s=0)}{L(\hat{n}_s, \hat{\gamma})}\right],
\end{align*} which is the logarithmic ratio between the likelihood of the background-only hypothesis $L(n_s=0)$ and the best-fit likelihood $L(\hat{n}_s, \hat{\gamma})$, where $\hat{n}_s$ and $\hat{\gamma}$ are the fitted values that maximize the likelihood~\cite{PS_analysis_methods}.
The background TS distribution is obtained with no injected signal by generating pseudo-experiments with randomized right ascensions. This background distribution is then fitted with a $\chi^2$ distribution, as motivated by Wilks’ theorem~\cite{Wilks1938}. The 90\,\% confidence level (C.L.) sensitivity is defined as the injected flux that yields a TS value exceeding the median of the background distribution in 90\,\% of the pseudo-experiments.
Similarly, the $5\sigma$ discovery potential is defined as the signal flux required for 50\,\% of trials to exceed the $5\sigma$ threshold of the background TS distribution.

The ELOWEN sensitivity, shown in~\Cref{sec:Results_transients} for comparison with the IceCube Upgrade transient sensitivity, is calculated under the assumption of a Poisson-distributed background of 20\,mHz, using a Bayesian approach~\cite{IceCube:ELOWEN_O4}.

\subsection{\label{sec:Framework_overview}Analyses Overview}

We perform three analyses to evaluate IceCube Upgrade sensitivities: one targeting time-dependent transients and two time-integrated searches targeted towards NGC 1068 and the GP. While all analyses use the same datasets introduced in~\Cref{sec:datasets} and follow the sensitivity calculation described in~\Cref{sec:Framework_likelihoods}, some incorporate a specific set of event cuts adopted to ensure robustness, as well as their own observational assumptions.

For the transient analysis, no additional event cuts are applied to the datasets, as the short observation timescale serves as the primary discriminator against background, suppressing artifacts from low-quality events deviating from the expected PDFs.
The analysis assumes that the burst position in the sky and the burst onset time are known, as is typical in searches for neutrino counterparts to sources detected via other messengers.
Accordingly, the sensitivity in this case corresponds to the flux level of a single, known burst that is detectable by the IceCube Upgrade.

In contrast to the transient analysis, time-integrated analyses, which rely purely on spatial and energy information over long livetimes, are sensitive to reconstruction artifacts from low-quality events. The baseline GRECO sample, in particular, contains a population of poorly reconstructed events that introduce significant distortions in the background test statistic distributions, making a robust time-integrated analysis infeasible without filtering.
Therefore, for the time-integrated analyses, the following specific quality cuts are applied:
\begin{itemize}
    \item GP: The IC93 sample is filtered with a simple energy cut on the reconstructed neutrino energy, $E_\text{reco}$, $2{\rm\,GeV}<E_\text{reco}<500{\rm\,GeV}$ (99.6\,\% signal retention for an $E^{-2.7}$ spectrum), while no additional cuts are applied to the GRECO datasets.
    \item NGC 1068: This point-source analysis requires more stringent filtering to ensure robust reconstruction. For the IC93 sample, the above energy cut is applied (99.0\,\% signal retention for an $E^{-3.4}$ spectrum) in addition to a cut on the reconstructed angular uncertainty $\sigma < 13^\circ$ (9.7\,\% signal retention), only keeping the best reconstructed events. Similarly, for the GRECO sample, a quality cut on angular error $< 7^\circ$ is applied (3.6\,\% signal retention) to select a high-purity, well-reconstructed event subsample.
\end{itemize}

The projections for the steady-state analyses compare two scenarios:
\begin{itemize}
    \item without Upgrade: Represents the sensitivity of the existing detector configuration. It uses the GRECO dataset, extrapolated forward in time for the subsequent years of operation.
    \item with Upgrade: Represents the future combined sensitivity. It consists of the extrapolated GRECO dataset (up to the IceCube Upgrade's integration foreseen in May 2026 - corresponding to 14 years) and the IC93 dataset, with the latter replacing GRECO after the IceCube Upgrade integration.
\end{itemize}
The GRECO extrapolation is performed by creating new data blocks via random sampling (without replacement) from the existing 11.6-year GRECO dataset. The number of sampled events is scaled to match the required future livetime. This data-driven method preserves the physical characteristics of the background, but relies on a single statistical realization, which can introduce small fluctuations in the sensitivity projections.

\section{\label{sec:Results}Projected sensitivities \& Discussion}

These analyses serve to probe the IceCube Upgrade's performance across different timescales (transient and steady-state) and source morphologies (point-source and extended). We present the results in two parts, reflecting the distinct analysis strategies required. 
First, we focus on transient sources, where the IceCube Upgrade's low-energy capabilities are most impactful. Since these analyses rely on short observation windows, we compare the sensitivities of the IC93 and GRECO datasets directly. 
Second, we examine steady-state sources, which rely on long-term data accumulation. For these time-independent studies, the projected sensitivity of the \textit{with Upgrade} scenario is compared to that of the \textit{without Upgrade} scenario across different detector livetimes, as defined in~\Cref{sec:Framework_overview}, to quantify the accumulated improvement over the existing detector.
In the figures, the neutrino flux is denoted by $dN/dE$ and the time-integrated flux is represented as $F(E)=(dN/dE)\Delta T$, where $\Delta T$ is the observation time.

\subsection{\label{sec:Results_transients}Transients}

\begin{figure}[tbp]
    \centering
    \includegraphics[width=0.95\linewidth]{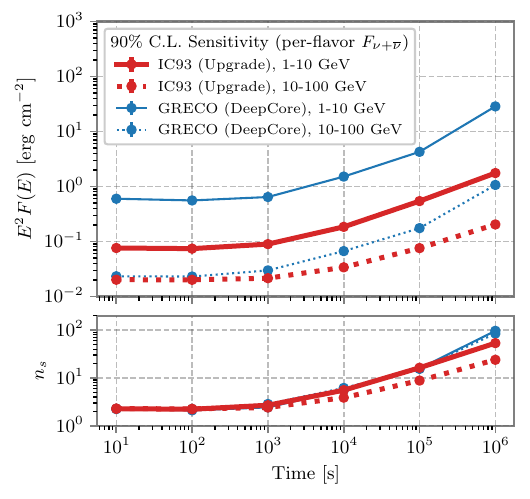}
    \caption{Sky-averaged per-flavor sensitivity of the IC93 sample and GRECO dataset as a function of the observation time window. The sensitivity is evaluated in two energy bins, 1--10\,GeV (solid lines) and 10--100\,GeV (dotted lines), assuming a source spectral index of $\gamma=2.0$.}
    \label{fig:transient_timewindow}
\end{figure}

In this section, we present the projected sensitivity of the IceCube Upgrade to short-timescale neutrino sources.
\Cref{fig:transient_timewindow} shows the sky-averaged IceCube Upgrade sensitivity as a function of the observation time window, compared to that of GRECO.
The sky average is computed over declinations ranging from $-60^\circ$ to $60^\circ$, sampled in $15^\circ$ increments, with a $\cos(\delta)$ weighting to properly account for the solid angle.
A significant improvement in the 1--10 GeV energy range is observed with the IceCube Upgrade, which can be attributed to the enhanced effective area in this range (see~\Cref{fig:datasets_effarea}).
The sensitivity remains independent of the observation duration up to a timescale of $\sim10^3$\,s, indicating the regime where the analysis is quasi background-free and signal-limited.
The number of atmospheric neutrino background events for a point-source observation with the IceCube Upgrade can be roughly estimated as
\begin{equation}
n_{\rm atm}^{\rm PS} \approx R_{\rm atm}^{\rm all}\Delta T\frac{2\pi(1-\cos\theta_{50})}{4\pi} \approx 0.4\left(\frac{\Delta T}{10^3\,\rm s}\right)
\label{eq:bg_rate_rep},
\end{equation}
where $R_{\rm atm}^{\rm all}=6.6$\,mHz is the sky-integrated atmospheric neutrino rate and $\theta_{50}\approx30^\circ$ denotes the median angular resolution in the GeV energy range (see \Cref{fig:datasets_angres}).
This estimate is consistent with the quasi background-free timescale of $\Delta T\sim10^3$\,s.
Beyond this timescale, background events begin to accumulate, resulting in a degradation in sensitivity.
An improvement is also evident in the 10--100\,GeV range in the background-dominated regime, driven by the enhanced background-rejection capability of the IceCube Upgrade.
Sensitivity on even longer timescales ($\gtrsim 1$\,yr) and across the entire spectrum, including applications to steady sources such as NGC~1068, is discussed in~\Cref{sec:Results_Steady}.

\begin{figure}[tbp]
    \centering
    \includegraphics[width=0.98\linewidth]{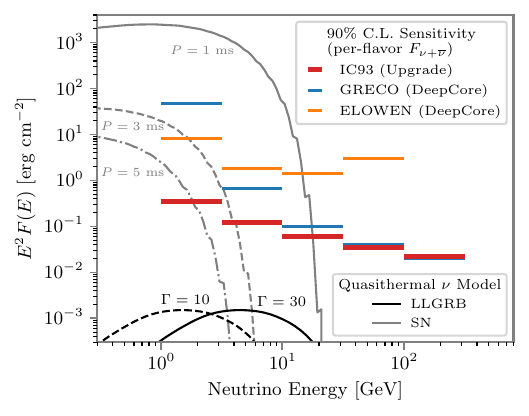}
    \caption{Sky-averaged per-flavor differential sensitivity of the IC93 sample and the existing DeepCore datasets.
    The signal is injected following a power-law spectrum with an index of $\gamma=2.0$.
    The observation time window is set to $\Delta T=1000$\,s.
    The black curves show the LLGRB quasi-thermal models with bulk Lorentz factor $\Gamma$ and a distance of 10\,Mpc~\cite{Murase_2022}.
    The gray curves indicate the SNe quasi-thermal models with the PNS rotational period $P$, magnetic field $B=10^{15}$\,G and a distance of 10\,kpc~\cite{Carpio2024}.}
    \label{fig:transient_diff_sens}
\end{figure}

\Cref{fig:transient_diff_sens} shows the differential sensitivity of the IC93 sample, together with those of the GRECO and ELOWEN datasets.
The signal is injected with a power-law spectrum with an index of $\gamma=2.0$. 
The observation time window is set to $\Delta T = 1000$\,s, which roughly corresponds to the quasi background-free observation timescale for the IC93 sample.
This timescale is longer than or comparable to the duration of quasi-thermal neutrino emission predicted from LLGRBs and SNe~\cite{Murase_2022, MuraseSN2014, Carpio2024}.
The black curves indicate the predicted quasi-thermal neutrino emission from LLGRBs with a bulk Lorentz factor of $\Gamma=10$ and $\Gamma=30$~\cite{Murase_2022}, assuming an isotropic-equivalent baryon kinetic energy of $\mathcal{E}_N=5\times10^{50}$\,erg and a source distance of 10\,Mpc.
The gray curves represent the expected quasi-thermal neutrino emission from SNe forming a PNS with the magnetic field of $B=10^{15}$\,G and the rotational period of $P=1$, 3, and 5\,ms, at a distance of 10\,kpc~\cite{Carpio2024}.
The IC93 sample will significantly improve the sensitivity at GeV energies, achieving sensitivities an order of magnitude better than the DeepCore datasets below 10\,GeV.

\begin{figure*}[!htbp]
  \centering
  \includegraphics[width=\textwidth]{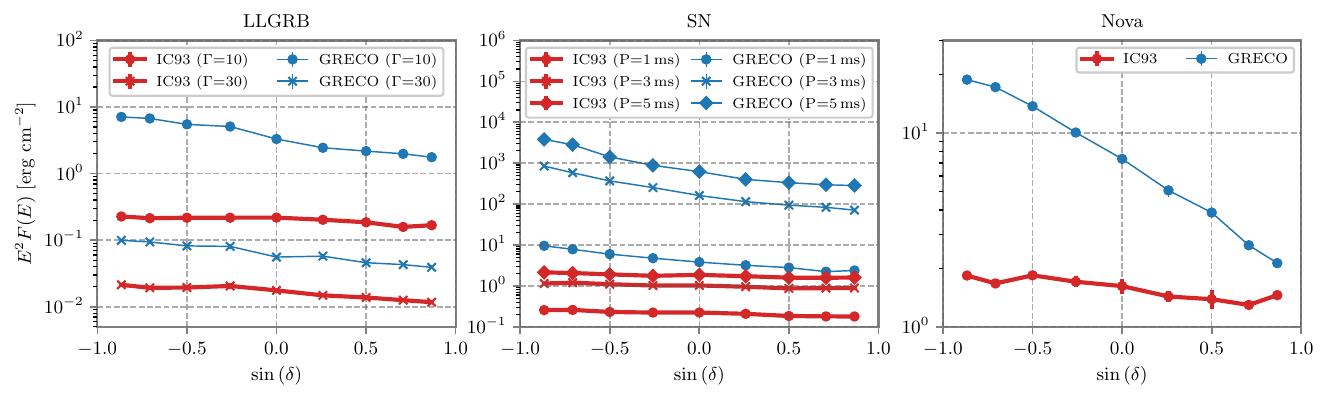}
  \caption{Per-flavor sensitivity of the IC93 sample to the transient neutrino emission models as a function of the source declination angle. The GRECO sensitivity is shown for a comparison. The time-integrated flux values are extracted at $E=1$\,GeV. (left) The LLGRB models with two different bulk Lorentz factors, 10 and 30, based on the templates introduced in~\cite{Murase_2022}. The time window is assumed to be $\Delta T=1000$\,s. (middle) The SN models with formation of a PNS with a magnetic field of $B=10^{15}$\,G and a rotational period of $P=1$, 3, and 5\,ms, extracted from~\cite{Carpio2024}, assuming a signal duration of $\Delta T=100$\,s. (right) The $\nu_\mu$ emission model expected from the nova RS~Oph, taken from~\cite{magic2022rsoph}. The signal duration is assumed to be $\Delta T=$10\,days.}
  \label{fig:transient_model_sens}
\end{figure*}

To quantify the expected improvement, the sensitivity to individual emission models is shown in~\Cref{fig:transient_model_sens}.
For LLGRBs with a bulk Lorentz factor of $\Gamma \sim 10$–30, the IC93 sample is expected to achieve roughly an order of magnitude better sensitivity than GRECO, as shown in the left panel of~\Cref{fig:transient_model_sens}.
This gain becomes more pronounced for GRBs with smaller $\Gamma$, whose spectra peak at lower energies.
The resulting sensitivity of $\sim0.1\,\rm erg\,cm^{-2}$ corresponds to a detectable distance on the Mpc scale.
LLGRBs within the IceCube Upgrade's reach would be extremely rare, given their estimated burst density of $\sim10^{-7}$–$10^{-6}\,\rm Mpc^{-3}\,yr^{-1}$~\cite{Liang_2007, Sun_2015}.
Nevertheless, the IceCube Upgrade is expected to significantly extend the sensitive coverage and provide the most stringent constraints on the population of general transients capable of emitting GeV neutrinos, as discussed later in this section.

One of the most promising transient classes for the IceCube Upgrade may be SNe that leave behind a PNS with a strong magnetic field and rapid rotation. Because their predicted spectra sharply drop in the 1–10\,GeV range, as illustrated in~\Cref{fig:transient_diff_sens}, the IC93 sample can significantly improve their detectability compared to GRECO.
This is shown in the middle panel of~\Cref{fig:transient_model_sens} for the PNS emission models with $B=10^{15}$\,G.
The IC93 sample enhances sensitivity to these PNS models by more than an order of magnitude relative to GRECO.
The IC93 sensitivity translates into a detectable distance of 40\,kpc (10\,kpc) for a PNS with a rotational period of $P = 3\,\rm ms~(5\,ms)$, whereas with GRECO it is only on the order of 1\,kpc.
These results suggest that Galactic SN events forming a PNS with a magnetic field of $B = 10^{15}$\,G and a rotational period $P \lesssim 5\,\rm ms$ are uniquely detectable with the IceCube Upgrade.
A PNS with more extreme parameters, such as $B = 10^{15}$\,G and $P = 1\,\rm ms$, may even be detectable on the Mpc scale, including in our neighboring galaxies.
We note, however, that the expected formation rate of rapidly rotating, strongly magnetized PNSs is anticipated to be lower than that of core-collapse SNe~\cite{Carpio2024}.

The IceCube Upgrade is also expected to improve sensitivity to Galactic novae, as their neutrino spectra are predicted to be soft and to peak around GeV energies.
We adopt the neutrino emission model presented in~\cite{magic2022rsoph}, which was developed from gamma-ray observations of the 2021 outburst of RS~Ophiuchi, the only nova detected in both the high- and very-high-energy gamma-ray bands~\cite{Fermi2022RSOph, magic2022rsoph, HESS2022RSOph, LST2025RSOph}.
The projected sensitivities for the IC93 sample and GRECO, assuming a signal duration of $\Delta T = 10$\,days, are shown in the right panel of~\Cref{fig:transient_model_sens}. 
Unlike the short-duration, background-free searches for GRBs and SNe, the 10-day time window adopted for the nova analysis falls within the background-limited regime, as it is significantly longer than the quasi background-free timescale estimated in \Cref{eq:bg_rate_rep}.
The sensitivities presented in~\Cref{fig:transient_model_sens} (right) were calculated accordingly, fully accounting for the accumulation of atmospheric background over the 10-day period.
The estimated sensitivities suggest that the IC93 sample can enhance the detectability of Galactic novae, particularly in the southern sky which is complemented by the results for steady point sources in the following section.
The efficient rejection of the atmospheric muon background in the IC93 sample drives a significant sensitivity improvement in the southern sky, which is particularly valuable since Galactic novae are predominantly located in this hemisphere hosting the galactic center.

\begin{figure}[b]
    \centering
    \includegraphics[width=0.95\linewidth]{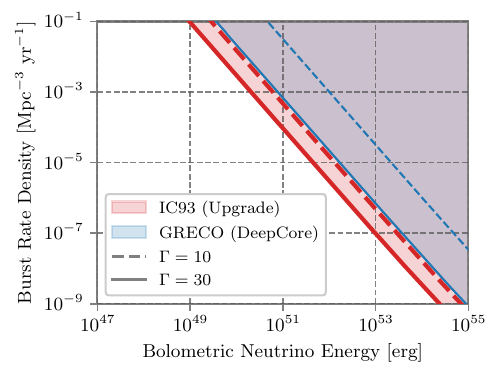}
    \caption{The shaded regions indicate burst populations that yield an expected detectable event rate exceeding 0.1 yr$^{-1}$ for IC93 and GRECO. The 90\,\% C.L. sensitivity for the GRB models shown in~\Cref{fig:transient_model_sens}, assuming $\Delta T=1000$\,s, is converted into the corresponding volume coverage as a function of the burst energy, and the event rate is computed from the volume and the assumed burst rate.}
    \label{fig:transient_param_space}
\end{figure}

Although detecting Galactic novae would remain challenging even with the IceCube Upgrade, a nearby and exceptionally bright nova would provide a unique opportunity to place stringent constraints on their neutrino emission or to detect a signal.
One promising candidate is T Coronae Borealis (T~CrB), which is predicted to experience an outburst within the next few years~\cite{Schaefer2023, Thwaites2025ICRC}, about 80 years after its last eruption in 1946.
When scaled by distance, T~CrB is expected to be about seven times brighter than RS~Oph, making it an excellent target for probing neutrino emission from novae~\cite{LST2025RSOph}.

Even if no specific transients, such as those mentioned above, are detected, the non-detection of neutrino bursts can still be used to place constraints on the general burst population, as demonstrated by the general transient search with DeepCore~\cite{IceCube:GRECO3yr}. \Cref{fig:transient_param_space} illustrates the population of GeV transients that can be probed with the IceCube Upgrade, extending significantly beyond the reach of GRECO. The LLGRB models are used as a proxy. The sensitivity shown in \Cref{fig:transient_model_sens} (left) is expressed in terms of a detectable distance, and the corresponding source population that yields more than 0.1 events yr$^{-1}$ within this distance is presented as a function of burst energy (approximated as 10\,\% of the baryon kinetic energy, $0.1\,\mathcal{E}_N$) and burst density. A uniform source distribution is assumed. Cosmological effects are not apparent, as the sensitivity range is much shorter than cosmological distances. The IceCube Upgrade is expected to provide the most stringent constraints on the transient population in the GeV energy range.
Although \Cref{fig:transient_param_space} cannot be directly compared to the results in~\cite{IceCube:GRECO3yr} due to different treatments of the source distance and assumptions on the spectral shape, it suggests that the IceCube Upgrade can extend DeepCore’s search to the GeV energy scale with a similar level of constraining power.

As demonstrated in this section, the IceCube Upgrade is expected to provide key sensitivity to transients in the $<100$\,GeV range in the coming decade.
Note that the sensitivity presented in this section is calculated for a single source.
Stacking observations from multiple sources can further enhance sensitivity, as is commonly done in IceCube analyses~\cite{GRECO_nova2023, GRECO_GRB2024}.

\subsection{\label{sec:Results_Steady}Steady-State Sources}

\begin{figure*}[tbp]
    \includegraphics[width=1\linewidth]
    {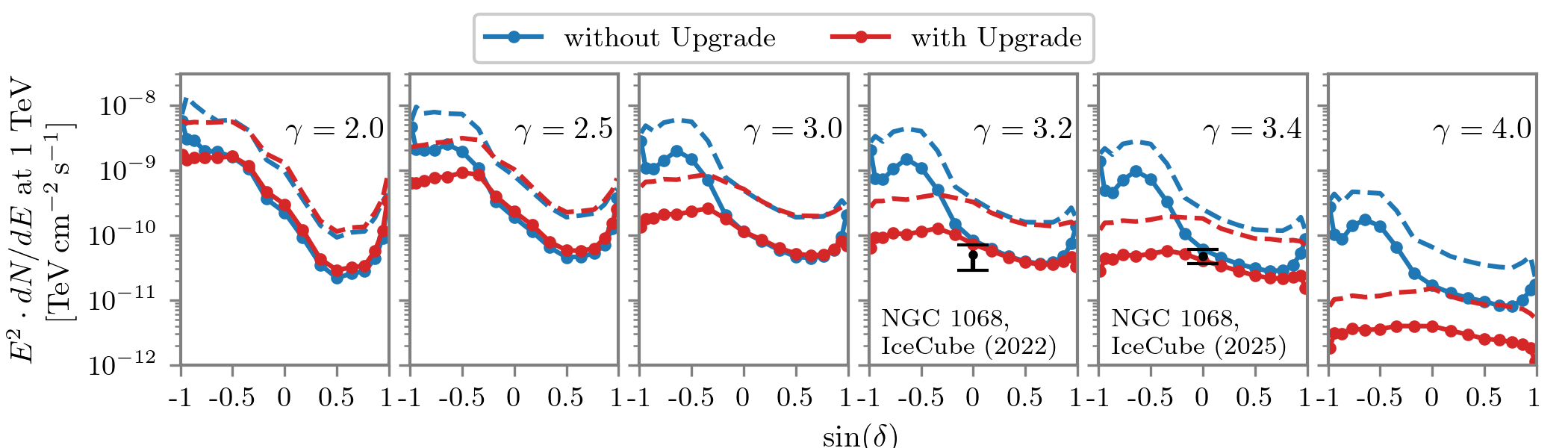}
    \caption{Projected 90\,\% C.L. sensitivity (solid) and $5\sigma$ discovery potential (dashed) as a function of source declination for various spectral indices. For spectral indices of $\gamma = 3.2$ and $\gamma = 3.4$, the best-fit flux from~\cite{IceCube:NGC1068,IceCube:NGC1068_FollupUp} is plotted in black for comparison.}
    \label{fig:ngc_decs}
\end{figure*}

For time-independent searches, detector data accumulates over many years of operation. We examine two benchmark cases: the point source NGC 1068 and the diffuse emission from the GP. These sources represent distinct physics cases: NGC 1068 is a soft-spectrum source ($\gamma \approx 3.2 - 3.4$)~\cite{IceCube:NGC1068, IceCube:NGC1068_FollupUp}, while the GP is expected to have a harder cosmic-ray-induced spectrum ($\gamma \approx 2.5 - 2.7$)~\cite{templates,FermiLAT:GP}. This contrast allows us to probe the energy-dependent performance of the IceCube Upgrade.
While these sources have been observed with high-energy neutrino datasets at TeV energies, the IceCube Upgrade provides an opportunity to probe their emission at GeV-scale energies.

The performance for point sources is summarized in~\Cref{fig:ngc_decs}, which shows the sensitivity as a function of declination across different spectral indices. A clear trend emerges: the IceCube Upgrade's improvement over the existing detector is most pronounced for soft spectra ($\gamma \gtrsim 3.0$).
In the northern sky ($\sin (\delta) > 0$), where the Earth provides a natural shield from atmospheric muons, the IceCube Upgrade's lower energy threshold improves sensitivity to the soft signal tail. Conversely, in the southern sky ($\sin (\delta) < 0$), improvements are achieved even for harder spectra despite the high atmospheric muon background. 
This enhancement is attributed to superior event reconstruction and more effective vetoing of the atmospheric muon background, enabled by the IceCube Upgrade’s denser instrumentation and the multi-PMT modules.
This southern sky enhancement is particularly valuable for searches for Galactic neutrino sources, such as supernova remnants or pulsar wind nebulae, as the bulk of the Galaxy is located in the southern hemisphere. It also benefits searches for extragalactic sources similar to NGC 1068 that are located in the south (e.g. Circinus Galaxy, NGC 7582, and ESO 138-1 as investigated in~\cite{NGC:OtherSources, IceCube:ESTES2026}).
For a source near the celestial equator, such as NGC\,1068, the sensitivity gain for its observed $\gamma=3.4$ spectrum is evident, however, it becomes even more substantial for softer spectra.

\begin{figure*}[t]
  \includegraphics[width=0.85\textwidth]{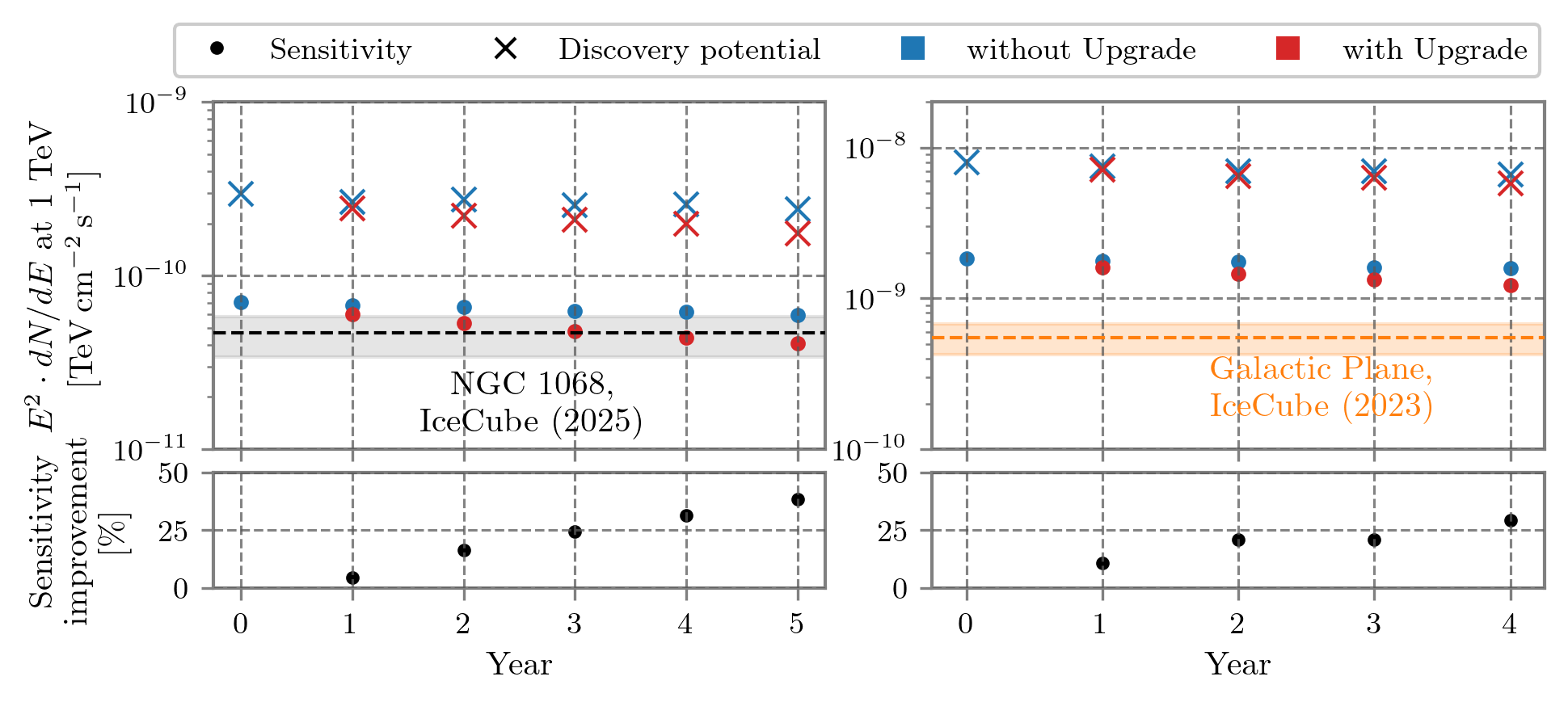}
    \caption{Temporal evolution of the projected sensitivity for the \textit{without Upgrade} (blue) and \textit{with Upgrade} (red) scenarios. (left) The point source NGC 1068 ($\gamma=3.4$) compared to the best-fit flux from~\cite{IceCube:NGC1068_FollupUp}. (right) The diffuse emission from the GP using the $\pi^0$ template ($\gamma=2.7$). }
    \label{fig:steady_time_evolution}
\end{figure*}

The temporal evolution of the sensitivity for both NGC 1068 and the GP is shown side-by-side in~\Cref{fig:steady_time_evolution}. For NGC 1068, the inclusion of IceCube Upgrade data results in a visibly steeper slope compared to the scenario without the IceCube Upgrade, yielding a relative sensitivity improvement of approximately 7.2\,\% per year for an $E^{-3.4}$ spectrum (reaching $\sim 31.5$\,\% after 4 years).
While modest compared to the transient case, this annual gain leads to an enhancement that accumulates over the livetime of the experiment.
A similar accumulating advantage is observed for the GP using the $\pi^0$ template~\cite{FermiLAT:GP}, assuming an $E^{-2.7}$ spectrum, where the IceCube Upgrade scenario achieves a $\sim30\,\%$ relative gain after 4 years.
The fluctuations observed in the GP projection result from the extrapolation method described in~\Cref{sec:Framework_overview}.

It is critical to note that the projections in this section are conservative with respect to the simulated energy range.
The IC93 dataset used here is strictly limited to events with $E_\text{true} \leq \SI{500}{GeV}$, whereas the \textit{without Upgrade} scenario uses only the GRECO dataset, which extends to $\sim 10$\,TeV.
Consequently, the IceCube Upgrade curves effectively discard all contributions from events with $E > 500$\,GeV past May 2026 that would still be detected—and better reconstructed—by the new instrumentation in reality. For soft sources like NGC 1068, this truncation has a minor impact (12.2\,\% annual improvement vs 7.2\,\%). For harder sources like the GP, the sensitivity is typically driven by the high-energy tail. When the cascades dataset ($500$\,GeV--$5$\,PeV), originally used for the observation of the GP as a neutrino source, is added to the GP analysis (not shown), the relative advantage of the low-energy IceCube Upgrade data diminishes if one strictly assumes a single, unbroken power law.
However, the inclusion of higher-statistic, low-energy IceCube Upgrade data is critical for determining how far the hard power law extends. While high-energy events drive the discovery potential for a standard power law, the IceCube Upgrade's low-energy threshold will play a decisive role in testing and constraining more complex emission models, such as broken power laws or spectral turnovers.
Although hard-spectrum searches are primarily driven by the main IceCube optical array's sensitivity to high-energy events, the IceCube Upgrade still delivers quantifiable sensitivity gains, complementing its specialized enhancements for soft spectra and low-energy phenomena.

\begin{figure}[tbp]
    \centering
    \includegraphics[width=0.95\linewidth]{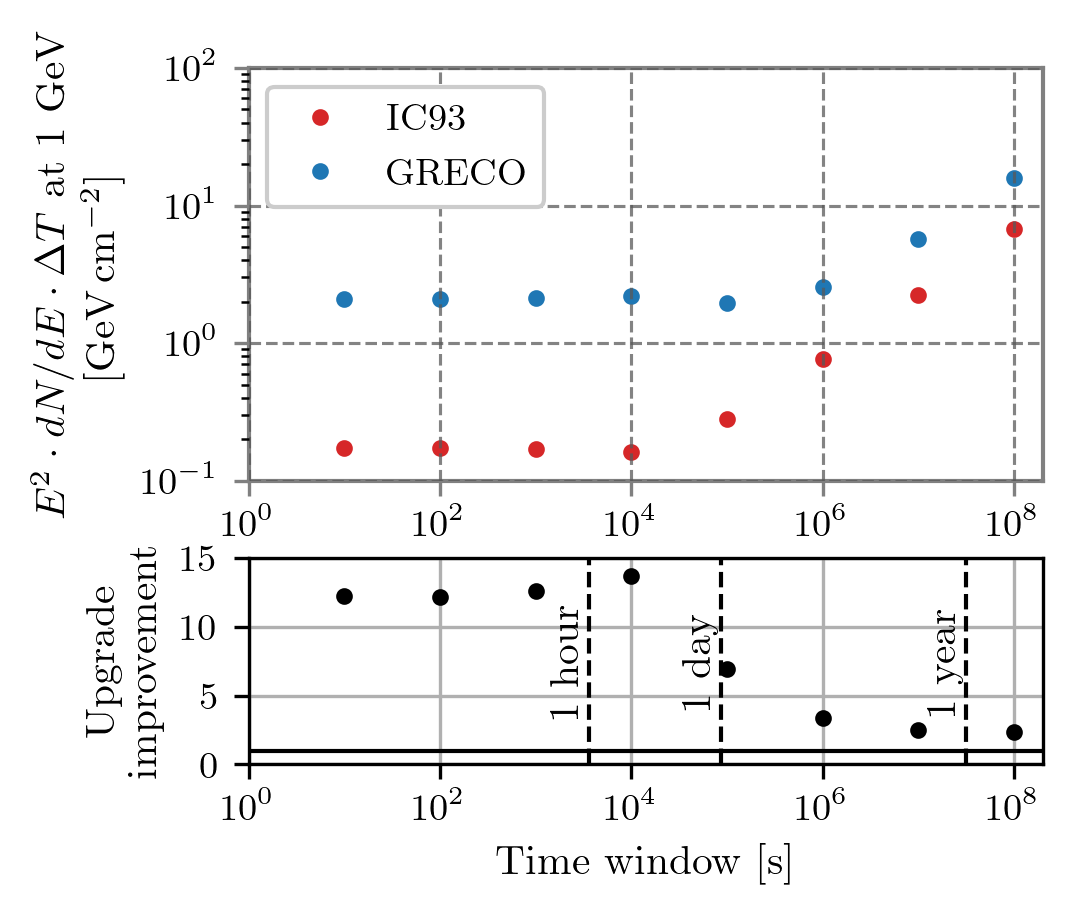}
        \caption{Sensitivity from direction of NGC 1068 at spectral index of 3.4 for both datasets across different time windows. The event cuts for the NGC 1068 analysis described in \Cref{sec:Framework_overview} are used. An increase corresponds to a transition from a background-free regime to a background-dominated regime.}
    \label{fig:timewindow_ngc}
\end{figure}

The moderate gains of $\sim 30\,\%$ for these steady-state sources stand in sharp contrast to the order-of-magnitude improvements seen in the transient analysis. This difference is purely driven by the background accumulation over the long 19-year livetime.
We visualize this transition in \Cref{fig:timewindow_ngc}, which shows a time-dependent analysis with the event cuts used in the steady-state NGC 1068 analysis, for the soft spectrum ($\gamma=3.4$) at $\delta=0^\circ$. This differs from the generic transient search presented in \Cref{fig:transient_timewindow}, which shows the sky-averaged performance for a source with $\gamma = 2$, without additional event cuts and binned for low energies only.
For short durations ($\Delta T \lesssim 10^4$\,s), the analysis remains in a signal-limited regime with negligible atmospheric background. Here, the sensitivity is driven by the Poissonian threshold (requiring $\sim 2.3$ signal events), and the IceCube Upgrade's larger low-energy effective area provides a massive advantage, mirroring the results from~\Cref{fig:transient_timewindow}. However, as the time window extends beyond $10^5$\,s, accumulating background events necessitate a larger signal flux for detection.
To suppress this background, the analysis effectively relies more on higher-energy events, a regime where the relative advantage of the IceCube Upgrade is smaller compared to the existing detector.
Consequently, the relative sensitivity improvement provided by the IceCube Upgrade diminishes, resulting in a convergence from the massive transient gains towards the modest but sustained improvements observed for steady-state sources across long livetimes.

\begin{figure}[t]
    \centering
    \includegraphics[width=0.95\linewidth]{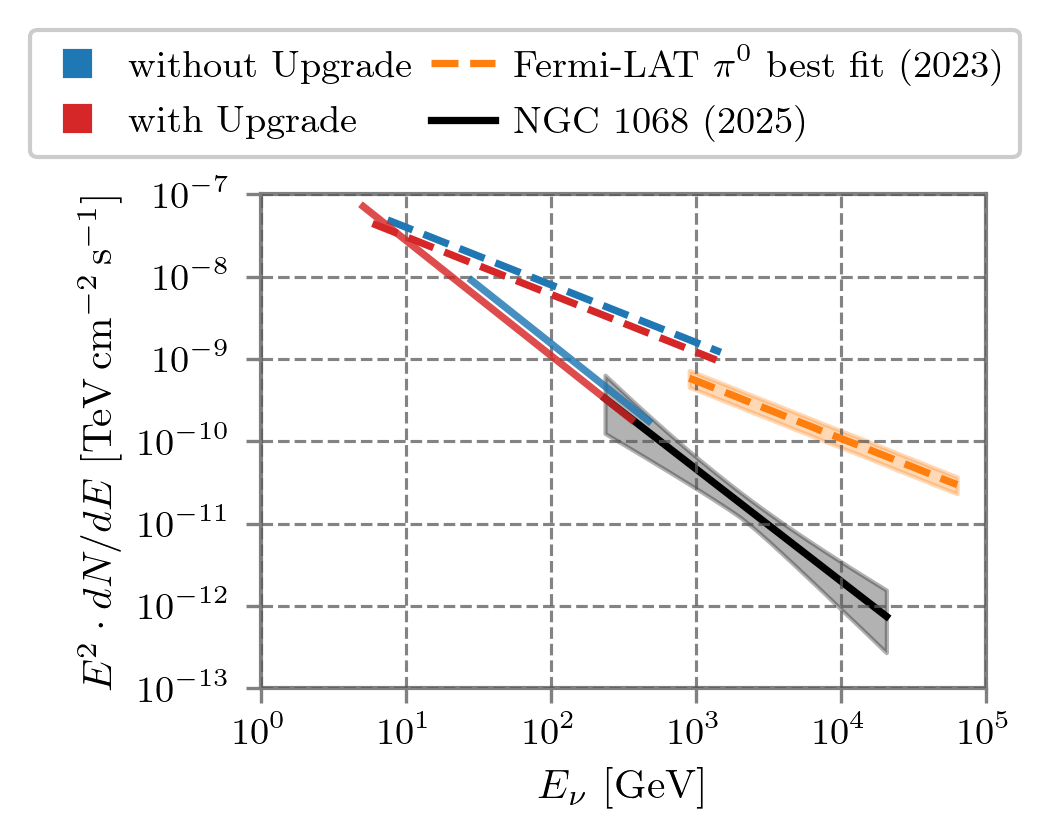}
    \caption{Comparison of projected IceCube Upgrade sensitivities (red) compared to the existing detector (blue) with published best-fit fluxes for NGC 1068 (black)~\cite{IceCube:NGC1068_FollupUp} and the GP ($\pi^0$ template, orange)~\cite{GP_Science_Paper}. 
    The GP flux is spatially integrated over the entire $\pi^0$ template.
    The sensitivity values correspond to the performance after four years of the IceCube Upgrade operation from~\Cref{fig:steady_time_evolution}, plotted as power laws over the sensitive energy range (central 90\,\% range in true neutrino energy) of the respective dataset configurations.}
    \label{fig:spectrum_summary}
\end{figure}

Finally, \Cref{fig:spectrum_summary} synthesizes the physics reach of the IceCube Upgrade for both examples of steady state sources in the context of the measured fluxes. Both scenarios (blue/red) for GP (dashed) and NGC 1068 (solid) represent the 4-year sensitivity projections derived from~\Cref{fig:steady_time_evolution}. The sensitivity is evaluated over the central 90\,\% interval of the expected signal distribution in true neutrino energy for each configuration. The GP flux is spatially integrated over the $\pi^0$ template to allow a direct comparison with the NGC 1068 flux on a common scale. The figure explicitly highlights the extension of the high sensitivity coverage into the 10--100\,GeV region. 

For soft-spectrum sources like NGC 1068, the IceCube Upgrade is projected to extend the effective sensitive energy range down to $\mathcal{O}(\text{GeV})$. For this scenario, sensitivity actually reaches the level of the best-fit flux in the TeV regime, indicating that the IceCube Upgrade can provide relevant constraints on the low-energy extension of the measured TeV emission, assuming a simple spectral extrapolation to GeV energies. Although such a straightforward extension is not strongly motivated theoretically, the IceCube Upgrade allows a direct test of this scenario.
While the sensitive energy range for the GP does not extend as noticeably as for NGC 1068, the sensitivity gains from including the IceCube Upgrade are comparable. The jump in sensitivity between the IceCube Upgrade and the high-energy measurements arises because high-energy datasets, such as \textit{cascades}~\cite{DNN}, were not included in this calculation.

This visualization reinforces that, although the main array continues to drive high-energy discovery, the IceCube Upgrade provides a complementary window for observing the low-energy components of these astrophysical fluxes with improved sensitivity.

\section{\label{sec:IC91}Impact of the deployed detector configuration}

\begin{figure}[!bh]
    \centering
    \includegraphics[width=0.95\linewidth]{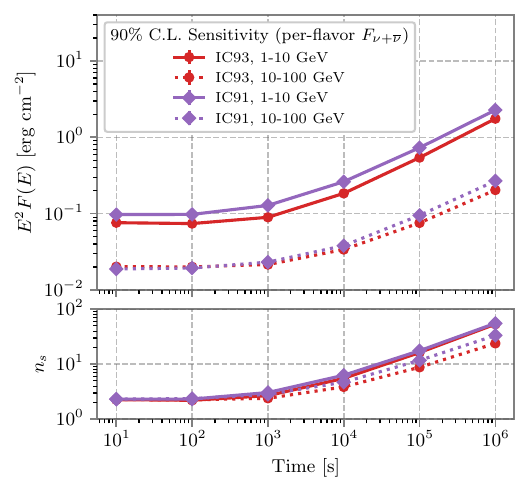}
    \caption{Sky-averaged per-flavor sensitivity of the IC91 simulation in comparison to that of the IC93 sample as a function of the observation time window. The sensitivity is evaluated in two energy bins, 1--10\,GeV (solid lines) and 10--100\,GeV (dotted lines), assuming a source spectral index of $\gamma=2.0$.}
    \label{fig:IC91vsIC93}
\end{figure}

The sensitivity results presented above are based on the originally planned seven-string detector configuration.
Since then, the IceCube Upgrade has been deployed with five strings, and approximately 90\,\% of the optical modules have been successfully commissioned. In this section, we evaluate the impact of the deployed configuration, hereafter referred to as IC91, on the projected sensitivity.

To estimate the performance of IC91, a modified version of the IC93 simulation is constructed by removing all pulses associated with the two strings absent from the deployed configuration, as well as those associated with optical modules that have not yet been commissioned.
Each event is then reprocessed using the procedure described in \Cref{sec:datasets}, including the trigger, filtering, event cuts, and reconstruction steps. Note that the machine-learning models used for noise cleaning, event classification, and reconstruction are not retrained for the IC91 configuration, nor are the simple event-selection cuts re-optimized. Consequently, the performance estimate presented here is expected to be conservative. Finally, the sensitivity of IC91 is estimated using the framework described in \Cref{sec:Framework}

\Cref{fig:IC91vsIC93} shows the same transient-sensitivity comparison as in \Cref{fig:transient_timewindow}, but between IC93 and IC91 instead of IC93 and GRECO.
The largest impact is observed at the lowest energies across all timescales, with IC91 showing a sensitivity degradation of approximately 30\,\% relative to IC93 in the 1--10\,GeV energy range.
This reduction is mainly caused by the smaller effective area resulting from the reduced number of strings and optical modules.
While the effective area remains nearly unchanged in the 10--100 GeV energy range with IC91, a sensitivity degradation of approximately 20\,\% is observed at timescales of $10^5$--$10^6$\,s, likely reflecting poorer directional reconstruction.

For the NGC 1068 analysis, the projected five-year sensitivity improvement is reduced to $\sim$10\,\%, compared to $\sim$40\,\% with IC93.
For the GP, the projected improvement for a four-year exposure decreases from $\sim$30\,\% to $\sim$20\,\%.
Nevertheless, the sensitivity improvement remains substantial for point sources at southern declinations, as well as for sources with soft energy spectra, although the spectral index at which a significant improvement is observed shifts to higher values ($\gamma \gtrsim 3.4$) than for IC93.
In addition, the sensitivities obtained with the IC91 dataset retain the capability to extend its sensitivity to energies of order GeV for soft-spectrum sources such as NGC 1068.
Assuming $\gamma=3.4$, the energy threshold increases from 5\,GeV for IC93 to 7\,GeV for IC91, but remains substantially lower than the $\sim$30\,GeV threshold of GRECO.

While these estimates indicate a non-negligible impact of the reduced number of strings and optical modules on the projected sensitivities presented in this study, the IceCube Upgrade is expected to continue to provide the best sensitivity at GeV energies, thereby extending the reach of existing observations down to these energies.
As the current estimate does not include optimization of the event selection for the deployed configuration, room for further improvement in sensitivity remains.

\section{\label{sec:Conclusion}Conclusions and Outlook}

The IceCube Upgrade has been deployed during the polar season of 2025--2026 and is designed to enhance sensitivity to low-energy neutrinos in the GeV range as well as reduce systematic uncertainties through dedicated high-precision calibration. While first data is expected in 2026, the full integration of the new optical modules into the data acquisition and analysis chains will be a complex process. 
The astrophysical projections presented in this work, based on initial simulations, show that the IceCube Upgrade's impact depends on the source's emission timescale, energy spectrum as well as the detection channel.

The most significant performance boost is projected for transient searches. In the quasi background-free, signal-limited regime ($\Delta T \lesssim 10^3$ s), the IceCube Upgrade's increased effective area at GeV energies yields sensitivity improvements of up to an order of magnitude. This capability opens a new window for detecting low-energy transient phenomena, like LLGRBs, core-collapse SNe, and Galactic novae.

For steady-state sources, sensitivity gains accumulate over observation times. Though the relative improvement is small due to the extensive archival data from the existing detector,
a consistent sensitivity gain is projected over the coming years. We expect the most significant improvements for soft-spectrum sources and those located in the southern sky. The enhancement of the latter is largely driven by the advanced optical modules, which enable superior event reconstruction and more effective rejection of the atmospheric muon background. Furthermore, the IceCube Upgrade provides critical sensitivity in the 1--100\,GeV energy band, significantly extending the observation window of the detector down to energies previously inaccessible with sufficient statistics.

These findings align with one of the IceCube Upgrade's primary design goals: enhancing performance at lower energies ($\lesssim 100$\,GeV)~\cite{IceCube:Upgrade}.
Another key objective of the IceCube Upgrade is to improve understanding of the surrounding glacial and refrozen borehole ice.
The new generation of calibration devices is expected to reduce dominant systematic uncertainties across the entire detector.
These improvements can be applied retrospectively to archival data, including higher-energy events.

The projections presented in this paper are based on the IC93 simulation, using the originally planned detector configuration rather than the deployed array, and therefore yield somewhat optimistic sensitivity projections.
Nevertheless, the estimated impact of the deployed detector configuration suggests that the overall scientific picture is not expected to change significantly.
The IC93 MC dataset also has known limitations in its energy range and background statistics.
Furthermore, the potential sensitivity gains arising from improved calibration and reduced systematic uncertainties are not yet accounted for.
These projections will be refined with future simulations and, in the near future, with data from the deployed array.

\begin{acknowledgements}
The authors gratefully acknowledge the support from the following agencies and institutions:
USA {\textendash} U.S. National Science Foundation-Office of Polar Programs,
U.S. National Science Foundation-Physics Division,
U.S. National Science Foundation-EPSCoR,
U.S. National Science Foundation-Office of Advanced Cyberinfrastructure,
Wisconsin Alumni Research Foundation,
Center for High Throughput Computing (CHTC) at the University of Wisconsin{\textendash}Madison,
Open Science Grid (OSG),
Partnership to Advance Throughput Computing (PATh),
Advanced Cyberinfrastructure Coordination Ecosystem: Services {\&} Support (ACCESS),
Frontera and Ranch computing project at the Texas Advanced Computing Center,
U.S. Department of Energy-National Energy Research Scientific Computing Center,
Particle astrophysics research computing center at the University of Maryland,
Michigan State University,
Astroparticle physics computational facility at Marquette University,
NVIDIA Corporation,
and Google Cloud Platform;
Belgium {\textendash} Funds for Scientific Research (FRS-FNRS and FWO),
FWO Odysseus and Big Science programmes,
and Belgian Federal Science Policy Office (Belspo);
Germany {\textendash} Bundesministerium f{\"u}r Forschung, Technologie und Raumfahrt (BMFTR),
Deutsche Forschungsgemeinschaft (DFG),
Helmholtz Alliance for Astroparticle Physics (HAP),
Initiative and Networking Fund of the Helmholtz Association,
Deutsches Elektronen Synchrotron (DESY),
and High Performance Computing cluster of the RWTH Aachen;
Sweden {\textendash} Swedish Research Council,
Swedish Polar Research Secretariat,
National Academic Infrastructure for Supercomputing in Sweden (NAISS),
and Knut and Alice Wallenberg Foundation;
European Union {\textendash} EGI Advanced Computing for research;
Australia {\textendash} Australian Research Council;
Canada {\textendash} Natural Sciences and Engineering Research Council of Canada,
Calcul Qu{\'e}bec, Compute Ontario, Canada Foundation for Innovation, WestGrid, and Digital Research Alliance of Canada;
Denmark {\textendash} Villum Fonden, Carlsberg Foundation, and European Commission;
New Zealand {\textendash} Marsden Fund;
Japan {\textendash} Japan Society for Promotion of Science (JSPS), Ministry of Education, Culture, Sports, Science and Technology (MEXT), and Institute for Global Prominent Research (IGPR) of Chiba University;
Korea {\textendash} National Research Foundation of Korea (NRF);
Switzerland {\textendash} Swiss National Science Foundation (SNSF).
\end{acknowledgements}

\vspace{-\baselineskip}

\appendix

\section{Signal-subtraction Method} \label{app:sig_sub}
In~\Cref{sec:Framework_likelihoods}, the unbinned maximum-likelihood method was described as a standard approach used in the search for point or extended neutrino sources.
The equations introduced in \Cref{sec:Framework_likelihoods} need to be modified for the search for the emission of neutrinos from the GP, since it covers a large area in declination. As a consequence, randomizing the right ascension does not guarantee background-only pseudo-experiments. This requires a modification of the likelihood function, known as \textit{signal-subtraction method}, which results in the following likelihood function:
\begin{align*}
    L(n_s,\gamma) = \prod_{i=1}^N \left[ \frac{n_s}{N} S_i + \tilde{D}_i - \frac{n_s}{N}\tilde{S}_i \right]
\end{align*} with 
\begin{align*}
    \tilde{D}_i = \left( 1- \frac{n_s}{N}\right) B_i + \frac{n_s}{N} \tilde{S}_i.
\end{align*} 
The probability density function (PDF) $\tilde{D}_i$ consists of the background PDF $B_i$ and the signal PDF $\tilde{S}_i$, where $\tilde{S}_i$ is the average of the right ascension~\cite{Signal_subtraction}. The likelihood for multiple datasets then changes to: 
\begin{align*}
    L(n_s,\gamma) = \prod_j^M \prod_{i \in j}^N \left[\frac{n_s^j}{N^j} S_i^j + \tilde{D}_i^j -\frac{n_s^j}{N^j} \tilde{S}_i^j \right].
\end{align*}
In addition, when analyzing the GP, the spectral index $\gamma$ was no longer treated as a free parameter but was fixed to the value corresponding to the template. 

\bibliography{sources}

@PREAMBLE{
 "\providecommand{\noopsort}[1]{}" 
 # "\providecommand{\singleletter}[1]{#1}%" 
}

@article{IceCube:QuesoPaper,
  author = {{IceCube Collaboration}},
  title = {{Physics potential of the IceCube Upgrade for atmospheric neutrino oscillations}},
  journal = {Phys. Rev. D},
  volume = {113},
  issue = {7},
  pages = {072009},
  numpages = {21},
  year = {2026},
  month = {Apr},
  publisher = {APS},
  doi = {10.1103/nnjw-jp1n},
  url = {https://doi.org/10.1103/nnjw-jp1n}
}

@article{IceCube:QuesoPaper_ICRC,
  author = {Eller,  Philipp and DeHolton,  Kayla Leonard and Weldert,  Jan and Ørsøe,  Rasmus},
  title = {{Sensitivity of the IceCube Upgrade to atmospheric neutrino oscillations}},
  journal = {PoS},
  collaboration = {IceCube},
  volume  = {ICRC2023},
  pages   = {1036},
  year    = {2023},
  doi     = {10.22323/1.444.1036},
  url = {https://pos.sissa.it/444/1036/}
}

@article{Abbasi2024,
  title = {{Search for 10–1000 GeV neutrinos from gamma-ray bursts with IceCube}},
  volume = {964},
  ISSN = {1538-4357},
  url = {http://dx.doi.org/10.3847/1538-4357/ad220b},
  DOI = {10.3847/1538-4357/ad220b},
  number = {2},
  journal = {ApJ},
  publisher = {American Astronomical Society},
  author = {{IceCube Collaboration}},
  year = {2024},
  month = mar,
  pages = {126}
}

@article{DNN,
doi = {10.1088/1748-0221/16/07/P07041},
url = {https://doi.org/10.1088/1748-0221/16/07/P07041},
year = {2021},
month = {jul},
publisher = {IOP Publishing},
volume = {16},
number = {07},
pages = {P07041},
author = {{IceCube Collaboration}},
title = {{A convolutional neural network based cascade reconstruction for the IceCube Neutrino Observatory}},
journal = {{JINST}},
}

@article{IceCube:IceCube,
  title = {{The IceCube Neutrino Observatory: Instrumentation and online systems}},
  volume = {12},
  ISSN = {1748-0221},
  url = {http://dx.doi.org/10.1088/1748-0221/12/03/P03012},
  DOI = {10.1088/1748-0221/12/03/p03012},
  number = {03},
  journal = {{JINST}},
  publisher = {IOP Publishing},
  author = {{IceCube Collaboration}},
  year = {2017},
  month = mar,
  pages = {P03012–P03012}
}

@article{Theory:Cherenkov,
 ISSN = {00368504, 20477163},
 URL = {http://www.jstor.org/stable/43425324},
 author = {R. E. Jennings},
 journal = {Science Progress (1933- )},
 number = {199},
 pages = {364--375},
 publisher = {Temporary Publisher},
 title = {{Čerenkov radiation}},
 urldate = {2025-10-30},
 volume = {50},
 year = {1962}
}

@article{IceCube:CrossSections,
  title = {Measurement of the high-energy all-flavor neutrino-nucleon cross section with IceCube},
  author = {{IceCube Collaboration}},
  journal = {Phys. Rev. D},
  volume = {104},
  issue = {2},
  pages = {022001},
  numpages = {11},
  year = {2021},
  month = {Jul},
  publisher = {American Physical Society},
  doi = {10.1103/PhysRevD.104.022001},
  url = {https://link.aps.org/doi/10.1103/PhysRevD.104.022001}
}

@article{IceCube:PMTPaper,
title = {{Calibration and characterization of the IceCube photomultiplier tube}},
journal = {NIM-A},
volume = {618},
number = {1},
pages = {139-152},
year = {2010},
issn = {0168-9002},
doi = {https://doi.org/10.1016/j.nima.2010.03.102},
url = {https://www.sciencedirect.com/science/article/pii/S0168900210006662},
author = {{IceCube Collaboration}},
}

@article{IceCube:DOM,
  title = {{The IceCube data acquisition system: Signal capture,  digitization,  and timestamping}},
  volume = {601},
  ISSN = {0168-9002},
  url = {http://dx.doi.org/10.1016/j.nima.2009.01.001},
  DOI = {10.1016/j.nima.2009.01.001},
  number = {3},
  journal = {NIM-A},
  publisher = {Elsevier BV},
  author = {{IceCube Collaboration}},
  year = {2009},
  month = apr,
  pages = {294–316}
}

@article{IceCube:DEgg,
  title = {{D-Egg: A dual PMT optical module for IceCube}},
  volume = {18},
  ISSN = {1748-0221},
  url = {http://dx.doi.org/10.1088/1748-0221/18/04/P04014},
  DOI = {10.1088/1748-0221/18/04/p04014},
  number = {04},
  journal = {{JINST}},
  publisher = {IOP Publishing},
  author = {{IceCube Collaboration}},
  year = {2023},
  month = apr,
  pages = {P04014}
}

@article{IceCube:mDOM,
      title={{Design and performance of the multi-PMT optical module for IceCube Upgrade}}, 
      author={Classen, Lew and others},
      collaboration = {IceCube},
      year={2021},
  journal = {PoS},
  volume  = {ICRC2023},
  pages   = {1070},
  doi     = {10.22323/1.395.1070 },
  url={https://pos.sissa.it/395/1070/pdf}, 
}

@article{IceCube:POCAM,
  author = "Khera, Nikhita and  Henningsen, Felix",
  title = "{POCAM in the IceCube Upgrade}",
  collaboration = {IceCube},
  journal = "PoS",
  year = 2021,
  volume = "ICRC2021",
  pages = "1049",
  url = {https://pos.sissa.it/395/1049/}
}

@inproceedings{IceCube:UpgradeCameraSystem,
  series = {NuFACT 2022},
  title = {{The camera system for the IceCube Upgrade}},
  url = {http://dx.doi.org/10.3390/psf2023008049},
  DOI = {10.3390/psf2023008049},
  booktitle = {NuFACT 2022},
  publisher = {MDPI},
  author = {Kang,  Woosik and others},
  year = {2023},
  month = sep,
  pages = {49},
  collection = {NuFACT 2022}
}

@article{IceCube:PencilBeam,
  title = {Advances in IceCube ice modelling  \& what to expect from the Upgrade},
  volume = {16},
  ISSN = {1748-0221},
  url = {http://dx.doi.org/10.1088/1748-0221/16/09/C09014},
  DOI = {10.1088/1748-0221/16/09/c09014},
  number = {09},
  journal = {Journal of Instrumentation},
  publisher = {IOP Publishing},
  author = {Rongen,  M. and Chirkin,  D.},
  collaboration = {IceCube},
  year = {2021},
  month = Sept,
  pages = {C09014}
}

@misc{Icecube:Gen2TDR,
    author       = {{IceCube-Gen2 Collaboration}},
    title        = {{IceCube-Gen2 technical design report}},
    year         = {2024},
    howpublished = {\url{https://icecube-gen2.wisc.edu/science/publications/tdr/}},
    note         = {Accessed: 2025-10-31},
}

@article{IceCube:NGC1068,
  title = {{Evidence for neutrino emission from the nearby active galaxy NGC 1068}},
  volume = {378},
  ISSN = {1095-9203},
  url = {http://dx.doi.org/10.1126/science.abg3395},
  DOI = {10.1126/science.abg3395},
  number = {6619},
  journal = {Science},
  publisher = {American Association for the Advancement of Science (AAAS)},
  author = {{IceCube Collaboration}},
  year = {2022},
  month = nov,
  pages = {538–543}
}

@article{IceCube:NGC1068_FollupUp,
doi = {10.3847/2041-8213/ae4aad},
url = {https://doi.org/10.3847/2041-8213/ae4aad},
year = {2026},
month = {mar},
publisher = {The American Astronomical Society},
volume = {1000},
number = {1},
pages = {L26},
author = {{IceCube Collaboration}},
title = {{Evidence for neutrino emission from X-ray-bright active galactic nuclei with IceCube}},
journal = {ApJL}
}

@article{IceCube:Upgrade,
  author  = {Ishihara, Aya},
  title   = {{The IceCube Upgrade -- design and science goals}},
  journal = {PoS},
  volume  = {ICRC2019},
  collaboration = {IceCube},
  pages   = {1031},
  year    = {2019},
  doi     = {10.22323/1.358.1031},
url={https://pos.sissa.it/358/1031/pdf}
}

@article{FermiLAT:GP,
  title = {{Fermi-LAT observations of the diffuse $\gamma$-ray emission: Implications for cosmic ray and the interstellar medium}},
  volume = {750},
  ISSN = {1538-4357},
  url = {http://dx.doi.org/10.1088/0004-637X/750/1/3},
  DOI = {10.1088/0004-637x/750/1/3},
  number = {1},
  journal = {ApJ},
  publisher = {American Astronomical Society},
  author = {Ackermann, M. and others},
  year = {2012},
  month = apr,
  pages = {3}
}

@article{GP_Science_Paper,
  title = {{Observation of high-energy neutrinos from the Galactic plane}},
  volume = {380},
  ISSN = {1095-9203},
  url = {http://dx.doi.org/10.1126/science.adc9818},
  DOI = {10.1126/science.adc9818},
  number = {6652},
  journal = {Science},
  publisher = {American Association for the Advancement of Science (AAAS)},
  author = {{IceCube Collaboration}},
  year = {2023},
  month = jun,
  pages = {1338–1343}
}

@article{PS_analysis_methods,
  title = {{Methods for point source analysis in high energy neutrino telescopes}},
  volume = {29},
  ISSN = {0927-6505},
  url = {http://dx.doi.org/10.1016/j.astropartphys.2008.02.007},
  DOI = {10.1016/j.astropartphys.2008.02.007},
  number = {4},
  journal = {Astropart. Phys.},
  publisher = {Elsevier BV},
  author = {{J. Braun et al.}},
  year = {2008},
  month = may,
  pages = {299–305}
}

@article{PS_analysis_multiple_datasets,
  title = {{Search for time-independent neutrino emission from astrophysical sources with 3 years of IceCube data}},
  volume = {779},
  ISSN = {1538-4357},
  url = {http://dx.doi.org/10.1088/0004-637X/779/2/132},
  DOI = {10.1088/0004-637x/779/2/132},
  number = {2},
  journal = {ApJ},
  publisher = {American Astronomical Society},
  author = {{IceCube Collaboration}},
  year = {2013},
  month = dec,
  pages = {132}
}

@article{Signal_subtraction,
  title = {{Constraints on galactic neutrino emission with seven years of IceCube data}},
  volume = {849},
  ISSN = {1538-4357},
  url = {http://dx.doi.org/10.3847/1538-4357/aa8dfb},
  DOI = {10.3847/1538-4357/aa8dfb},
  number = {1},
  journal = {ApJ},
  publisher = {American Astronomical Society},
  author = {{IceCube Collaboration}},
  year = {2017},
  month = oct,
  pages = {67}
}

@article{templates,
   title={{The gamma-ray and neutrino sky: A consistent picture of Fermi-LAT, Milagro, and IceCube results}},
   volume={815},
   ISSN={2041-8213},
   url={http://dx.doi.org/10.1088/2041-8205/815/2/L25},
   DOI={10.1088/2041-8205/815/2/l25},
   number={2},
   journal={ApJ},
   publisher={American Astronomical Society},
   author={Gaggero, Daniele and others},
   year={2015},
   month=dec, pages={L25} }

@inproceedings{Stuttard2020,
  series = {NuFact2019},
  title = {{Neutrino oscillations and PMNS unitarity with IceCube/DeepCore and the IceCube Upgrade}},
  url = {http://dx.doi.org/10.22323/1.369.0099},
  DOI = {10.22323/1.369.0099},
  booktitle = {PoS(NuFact2019)},
  publisher = {Sissa Medialab},
  author = {Stuttard,  Tom},
  year = {2020},
  month = jun,
  pages = {099},
  collection = {NuFact2019}
}

@article{Murase_2022,
    doi = {10.3847/2041-8213/aca3ae},
    url = {https://doi.org/10.3847/2041-8213/aca3ae},
    year = {2022},
    month = {dec},
    publisher = {The American Astronomical Society},
    volume = {941},
    number = {1},
    pages = {L10},
    author = {Murase, Kohta and others},
    title = {{Neutrinos from the brightest gamma-ray burst?}},
    journal = {ApJL},
}

@article{Carpio2024,
  title = {{Quasithermal GeV neutrinos from neutron-loaded magnetized outflows in core-collapse supernovae: Spectra and light curves}},
  author = {Carpio, Jose A. and others},
  journal = {Phys. Rev. D},
  volume = {110},
  issue = {8},
  pages = {083012},
  numpages = {12},
  year = {2024},
  month = {Oct},
  publisher = {American Physical Society},
  doi = {10.1103/PhysRevD.110.083012},
  url = {https://link.aps.org/doi/10.1103/PhysRevD.110.083012}
}

@article{magic2022rsoph,
	author = {Acciari, V. A. and others},
	date = {2022/06/01},
	doi = {10.1038/s41550-022-01640-z},
	id = {Acciari2022},
	isbn = {2397-3366},
	journal = {Nature Astronomy},
	number = {6},
	pages = {689--697},
	title = {{Proton acceleration in thermonuclear nova explosions revealed by gamma rays}},
	url = {https://doi.org/10.1038/s41550-022-01640-z},
	volume = {6},
	year = {2022}}

@article{Liang_2007,
    doi = {10.1086/517959},
    url = {https://dx.doi.org/10.1086/517959},
    year = {2007},
    month = {jun},
    publisher = {},
    volume = {662},
    number = {2},
    pages = {1111},
    author = {Liang, Enwei and others},
    title = {{Low-luminosity gamma-ray bursts as a unique population: Luminosity function, local rate, and beaming factor}},
    journal = {ApJ},
}

@article{Sun_2015,
    doi = {10.1088/0004-637X/812/1/33},
    url = {https://dx.doi.org/10.1088/0004-637X/812/1/33},
    year = {2015},
    month = {oct},
    publisher = {The American Astronomical Society},
    volume = {812},
    number = {1},
    pages = {33},
    author = {Sun, Hui and Zhang, Bing and Li, Zhuo},
    title = {{Extragalactic high-energy transients: Event rate densities and luminosity functions}},
    journal = {ApJ},
}

@ARTICLE{Schaefer2023,
    author = {Schaefer, Bradley E},
    title = {{The B \& V light curves for recurrent nova T CrB from 1842–2022, the unique pre- and post-eruption high-states, the complex period changes, and the upcoming eruption in 2025.5±1.3}},
    journal = {MNRAS},
    volume = {524},
    number = {2},
    pages = {3146-3165},
    year = {2023},
    month = {03},
    issn = {0035-8711},
    doi = {10.1093/mnras/stad735},
    url = {https://doi.org/10.1093/mnras/stad735}
}

@article{Thwaites2025ICRC,
  author = "Thwaites, Jessie  and  others",
  title = {{Search for GeV-PeV neutrinos from nova T Coronae Borealis with IceCube}},
  doi = "10.22323/1.501.1200",
  collaboration = {IceCube},
  journal = "PoS",
  year = 2025,
  volume = "ICRC2025",
  pages = "1200"
}

@article{BECHERINI2006,
    title = {{A parameterisation of single and multiple muons in the deep water or ice}},
    journal = {Astropart. Phys.},
    volume = {25},
    number = {1},
    pages = {1-13},
    year = {2006},
    issn = {0927-6505},
    doi = {https://doi.org/10.1016/j.astropartphys.2005.10.005},
    url = {https://www.sciencedirect.com/science/article/pii/S092765050500157X},
    author = {Becherini, Y. and others},
}

@article{KOEHNE2013,
    title = {{PROPOSAL: A tool for propagation of charged leptons}},
    journal = {CPC},
    volume = {184},
    number = {9},
    pages = {2070-2090},
    year = {2013},
    issn = {0010-4655},
    doi = {https://doi.org/10.1016/j.cpc.2013.04.001},
    url = {https://www.sciencedirect.com/science/article/pii/S0010465513001355},
    author = {J.-H. Koehne and others},
}

@article{AGOSTINELLI2003,
    title = {{Geant4—a simulation toolkit}},
    journal = {NIM-A},
    volume = {506},
    number = {3},
    pages = {250-303},
    year = {2003},
    issn = {0168-9002},
    doi = {https://doi.org/10.1016/S0168-9002(03)01368-8},
    url = {https://www.sciencedirect.com/science/article/pii/S0168900203013688},
    author = {S. Agostinelli and others},
}

@article{ANDREOPOULOS2010,
    title = {{The GENIE neutrino Monte Carlo generator}},
    journal = {NIM-A},
    volume = {614},
    number = {1},
    pages = {87-104},
    year = {2010},
    issn = {0168-9002},
    doi = {https://doi.org/10.1016/j.nima.2009.12.009},
    url = {https://www.sciencedirect.com/science/article/pii/S0168900209023043},
    author = {C. Andreopoulos and others},
}

@ARTICLE{ANDREOPOULOS2015,
       author = {{Andreopoulos}, Costas and others},
        title = {{The GENIE neutrino Monte Carlo generator: Physics and user manual}},
      journal = {arXiv e-prints},
         year = 2015,
        month = oct,
          eid = {arXiv:1510.05494},
        pages = {arXiv:1510.05494},
          doi = {10.48550/arXiv.1510.05494},
}

@article{clsim,
  author = {{Chirkin}, Dmitry and others},
  journal={eScience 2019}, 
  title={{Photon propagation using GPUs by the IceCube Neutrino Observatory}}, 
  year={2019},
  volume={},
  number={},
  pages={388-393},
  doi={10.1109/eScience.2019.00050}
}

@article{Sogaard2023,
    doi = {10.21105/joss.04971},
    url = {https://doi.org/10.21105/joss.04971},
    year = {2023},
    publisher = {JOSS},
    volume = {8},
    number = {85},
    pages = {4971},
    author = {Andreas Søgaard and others},
    title = {{GraphNeT: Graph neural networks for neutrino telescope event reconstruction}},
    journal = {Journal of Open Source Software}
}

@article{Murase2013,
  title = {{Subphotospheric neutrinos from gamma-ray bursts: The role of neutrons}},
  author = {Murase, Kohta and Kashiyama, Kazumi and M\'esz\'aros, Peter},
  journal = {Phys. Rev. Lett.},
  volume = {111},
  issue = {13},
  pages = {131102},
  numpages = {5},
  year = {2013},
  month = {Sep},
  publisher = {American Physical Society},
  doi = {10.1103/PhysRevLett.111.131102},
  url = {https://link.aps.org/doi/10.1103/PhysRevLett.111.131102}
}

@article{GRECO_GRB2024,
    doi = {10.3847/1538-4357/ad220b},
    url = {https://doi.org/10.3847/1538-4357/ad220b},
    year = {2024},
    month = {mar},
    publisher = {The American Astronomical Society},
    volume = {964},
    number = {2},
    pages = {126},
    author = {{IceCube Collaboration}},
    title = {{Search for 10–1000 GeV neutrinos from gamma-ray bursts with IceCube}},
    journal = {ApJ},
}

@ARTICLE{ELOWEN2021,
  title = {{Search for GeV neutrino emission during intense gamma-ray solar flares with the IceCube Neutrino Observatory}},
  author = {{IceCube Collaboration}},
  journal = {Phys. Rev. D},
  volume = {103},
  issue = {10},
  pages = {102001},
  numpages = {12},
  year = {2021},
  month = {May},
  publisher = {American Physical Society},
  doi = {10.1103/PhysRevD.103.102001},
  url = {https://link.aps.org/doi/10.1103/PhysRevD.103.102001}
}

@article{GRECO_nova2023,
doi = {10.3847/1538-4357/acdc1b},
url = {https://doi.org/10.3847/1538-4357/acdc1b},
year = {2023},
month = {aug},
publisher = {The American Astronomical Society},
volume = {953},
number = {2},
pages = {160},
author = {{IceCube Collaboration}},
title = {{Search for sub-TeV neutrino emission from novae with IceCube-DeepCore}},
journal = {ApJ},

}

@article{GRECO_GW2024,
doi = {10.3847/1538-4357/aceefc},
url = {https://doi.org/10.3847/1538-4357/aceefc},
year = {2023},
month = {dec},
publisher = {The American Astronomical Society},
volume = {959},
number = {2},
pages = {96},
author = {{IceCube Collaboration}},
title = {{A search for IceCube sub-TeV neutrinos correlated with gravitational-wave events detected by LIGO/Virgo}},
journal = {ApJ},
}

@article{MuraseSN2014,
  title = {{Quasithermal neutrinos from rotating protoneutron stars born during core collapse of massive stars}},
  author = {Murase, Kohta and Dasgupta, Basudeb and Thompson, Todd A.},
  journal = {Phys. Rev. D},
  volume = {89},
  issue = {4},
  pages = {043012},
  numpages = {8},
  year = {2014},
  month = {Feb},
  publisher = {American Physical Society},
  doi = {10.1103/PhysRevD.89.043012},
  url = {https://link.aps.org/doi/10.1103/PhysRevD.89.043012}
}

@article{Fermi2022RSOph,
    doi = {10.3847/1538-4357/ac7eb7},
    url = {https://doi.org/10.3847/1538-4357/ac7eb7},
    year = {2022},
    month = {aug},
    publisher = {The American Astronomical Society},
    volume = {935},
    number = {1},
    pages = {44},
    author = {Cheung, C. C. and others},
    title = {{Fermi LAT gamma-ray detection of the recurrent nova RS Ophiuchi during its 2021 outburst}},
    journal = {ApJ},
}

@article{HESS2022RSOph,
    author = {F. Aharonian and others},
    title = {{Time-resolved hadronic particle acceleration in the recurrent nova RS Ophiuchi}},
    journal = {Science},
    volume = {376},
    number = {6588},
    pages = {77-80},
    year = {2022},
    doi = {10.1126/science.abn0567},
    URL = {https://www.science.org/doi/abs/10.1126/science.abn0567}
}

@article{LST2025RSOph,
	author = {{Abe, K.} and others},
	title = {{Detection of RS Oph with LST-1 and modelling of its HE/VHE gamma-ray emission}},
	DOI= "10.1051/0004-6361/202452447",
	url= "https://doi.org/10.1051/0004-6361/202452447",
	journal = {A\&A},
	year = 2025,
	volume = 695,
	pages = "A152",
}

@article{nu_osc_deepcore,
  title = {{Measurement of atmospheric neutrino oscillation parameters using convolutional neural networks with 9.3 years of data in IceCube DeepCore}},
  author = {{IceCube Collaboration}},
  collaboration = {IceCube Collaboration},
  journal = {Phys. Rev. Lett.},
  volume = {134},
  issue = {9},
  pages = {091801},
  numpages = {10},
  year = {2025},
  month = {Mar},
  publisher = {American Physical Society},
  doi = {10.1103/PhysRevLett.134.091801},
  url = {https://link.aps.org/doi/10.1103/PhysRevLett.134.091801}
}

@article{Abbasi2023,
  title = {Measurement of atmospheric neutrino mixing with improved IceCube DeepCore calibration and data processing},
  author = {{IceCube Collaboration}},
  collaboration = {IceCube Collaboration},
  journal = {Phys. Rev. D},
  volume = {108},
  issue = {1},
  pages = {012014},
  numpages = {36},
  year = {2023},
  month = {Jul},
  publisher = {American Physical Society},
  doi = {10.1103/PhysRevD.108.012014},
  url = {https://link.aps.org/doi/10.1103/PhysRevD.108.012014}
}

@article{Honda2015AtmoNu,
  title = {{Atmospheric neutrino flux calculation using the NRLMSISE-00 atmospheric model}},
  author = {Honda, M. and others},
  journal = {Phys. Rev. D},
  volume = {92},
  issue = {2},
  pages = {023004},
  numpages = {15},
  year = {2015},
  month = {Jul},
  publisher = {American Physical Society},
  doi = {10.1103/PhysRevD.92.023004},
  url = {https://link.aps.org/doi/10.1103/PhysRevD.92.023004}
}

@article{IceCube:Greco_intro,
  title = {{Measurement of atmospheric tau neutrino appearance with IceCube DeepCore}},
  volume = {99},
  ISSN = {2470-0029},
  url = {http://dx.doi.org/10.1103/PhysRevD.99.032007},
  DOI = {10.1103/physrevd.99.032007},
  number = {3},
  journal = {Phys. Rev. D},
  publisher = {American Physical Society (APS)},
  author = {{IceCube Collaboration}},
  year = {2019},
  month = feb 
}

@article{IceCube:ELOWEN_O4,
  title = {{Probing neutrino emission at GeV energies from compact binary mergers with the IceCube Neutrino Observatory}},
  author = {{IceCube Collaboration}},
  collaboration = {IceCube Collaboration},
  journal = {Phys. Rev. D},
  volume = {113},
  issue = {4},
  pages = {042003},
  numpages = {14},
  year = {2026},
  month = {Feb},
  publisher = {American Physical Society},
  doi = {10.1103/bkxc-4116},
  url = {https://link.aps.org/doi/10.1103/bkxc-4116}
}

@article{NGC:OtherSources,
  author = {Yu,  Shiqi and Kheirandish,  Ali and Liu,  Qinrui and Niederhausen,  Hans},
  title = {{Search for TeV neutrinos from Seyfert galaxies in the southern sky using starting track events in IceCube}},
  journal = {PoS},
  volume  = {ICRC2023},
  collaboration = {IceCube},
  pages   = {1533},
  year    = {2023},
  doi     = {10.22323/1.444.1533},
  url = {https://pos.sissa.it/444/1533/pdf}
}

@article{IceCube:DeepCoreDesign2012,
    title = {{The design and performance of IceCube DeepCore}},
    journal = {Astropart. Phys.},
    volume = {35},
    number = {10},
    pages = {615-624},
    year = {2012},
    issn = {0927-6505},
    doi = {https://doi.org/10.1016/j.astropartphys.2012.01.004},
    url = {https://www.sciencedirect.com/science/article/pii/S0927650512000254},
    author = {R. Abbasi and others},
}

@phdthesis{IceCube:MuonGun,
    author = "van Santen, Jakob",
    title = "{{Neutrino interactions in IceCube above 1 TeV: Constraints on atmospheric charmed-meson production and investigation of the astrophysical neutrino flux with 2 Years of IceCube data taken 2010--2012}}",
    school = "Wisconsin U., Madison",
    month = "11",
    year = "2014",
    url={https://asset.library.wisc.edu/1711.dl/GX7HPABPUJ5YO84/R/file-dddd0.pdf}
}

@ARTICLE{GNNModel,
  author={Scarselli, Franco and others},
  journal={IEEE Trans. Neural Netw.}, 
  title={{The graph neural network model}}, 
  year={2009},
  volume={20},
  number={1},
  pages={61-80},
  doi={10.1109/TNN.2008.2005605}
}

@article{GRBQTNu2000Bahcall,
  title = {{5-10 GeV neutrinos from gamma-ray burst fireballs}},
  author = {{Bahcall}, John N. and {M{\'e}sz{\'a}ros}, Peter},
  journal = {Phys. Rev. Lett.},
  volume = {85},
  issue = {7},
  pages = {1362--1365},
  numpages = {0},
  year = {2000},
  month = {Aug},
  publisher = {American Physical Society},
  doi = {10.1103/PhysRevLett.85.1362},
  url = {https://link.aps.org/doi/10.1103/PhysRevLett.85.1362}
}

@article{GRBQTNu2000Meszaros,
doi = {10.1086/312894},
url = {https://doi.org/10.1086/312894},
year = {2000},
month = {sep},
publisher = {},
volume = {541},
number = {1},
pages = {L5},
author = {Mészáros, P. and Rees, M. J.},
title = {{Multi-GeV neutrinos from internal dissipation in gamma-ray burst fireballs}},
journal = {ApJ},
}

@article{Nova:Review2020Chomiuk,
    author = "Chomiuk, Laura and Metzger, Brian D. and Shen, Ken J.",
    title = {{New insights into classical novae}},
    doi = "10.1146/annurev-astro-112420-114502",
    journal = "Ann. Rev. Astron. Astrophys.",
    volume = "59",
    pages = "391--444",
    year = "2021"
}

@article{SN:ReviewNu2018,
	author = {Tamborra, Irene and Murase, Kohta},
	date = {2018/01/23},
	doi = {10.1007/s11214-018-0468-7},
	id = {Tamborra2018},
	isbn = {1572-9672},
	journal = {Space Science Reviews},
	number = {1},
	pages = {31},
	title = {{Neutrinos from supernovae}},
	url = {https://doi.org/10.1007/s11214-018-0468-7},
	volume = {214},
	year = {2018}
}

@article{NGC1068Review,
	author = {Padovani, P. and others},
	date = {2024/09/01},
	doi = {10.1038/s41550-024-02339-z},
	id = {Padovani2024},
	isbn = {2397-3366},
	journal = {Nature Astronomy},
	number = {9},
	pages = {1077--1087},
	title = {{High-energy neutrinos from the vicinity of the supermassive black hole in NGC 1068}},
	url = {https://doi.org/10.1038/s41550-024-02339-z},
	volume = {8},
	year = {2024}
}

@article{IceCube:GRB2022,
doi = {10.3847/1538-4357/ac9785},
url = {https://doi.org/10.3847/1538-4357/ac9785},
year = {2022},
month = {nov},
publisher = {The American Astronomical Society},
volume = {939},
number = {2},
pages = {116},
author = {Abbasi, R. and others},
title = {{Searches for neutrinos from gamma-ray bursts using the IceCube Neutrino Observatory}},
journal = {ApJ},
}

@article{InteractSN_Murase2024,
  title = {{Interacting supernovae as high-energy multimessenger transients}},
  author = {Murase, Kohta},
  journal = {Phys. Rev. D},
  volume = {109},
  issue = {10},
  pages = {103020},
  numpages = {21},
  year = {2024},
  month = {May},
  publisher = {American Physical Society},
  doi = {10.1103/PhysRevD.109.103020},
  url = {https://link.aps.org/doi/10.1103/PhysRevD.109.103020}
}

@article{IceCube:GRECO3yr,
doi = {10.1088/1475-7516/2022/01/027},
url = {https://doi.org/10.1088/1475-7516/2022/01/027},
year = {2022},
month = {jan},
publisher = {IOP Publishing},
volume = {2022},
number = {01},
pages = {027},
author = {{IceCube Collaboration}},
title = {{First all-flavor search for transient neutrino emission using 3-years of IceCube DeepCore data}},
journal = {JCAP},
}

@article{IceCube:SN_2023,
doi = {10.3847/2041-8213/acd2c9},
url = {https://doi.org/10.3847/2041-8213/acd2c9},
year = {2023},
month = {may},
publisher = {The American Astronomical Society},
volume = {949},
number = {1},
pages = {L12},
author = {{IceCube Collaboration}},
title = {{Constraining high-energy neutrino emission from supernovae with IceCube}},
journal = {ApJL},
}

@article{IceCube:Diffuse2013,
author = {{IceCube Collaboration}},
title = {{Evidence for high-energy extraterrestrial neutrinos at the IceCube Detector}},
journal = {Science},
volume = {342},
number = {6161},
pages = {1242856},
year = {2013},
doi = {10.1126/science.1242856},
URL = {https://www.science.org/doi/abs/10.1126/science.1242856},
}

@article{GeVTransients_Sherman2025,
doi = {10.3847/1538-4357/adb716},
url = {https://doi.org/10.3847/1538-4357/adb716},
year = {2025},
month = {mar},
publisher = {The American Astronomical Society},
volume = {982},
number = {2},
pages = {94},
author = {Sherman, Angelina and others},
title = {{Prospects for observing astrophysical transients with gigaelectronvolt neutrinos}},
journal = {ApJ},
}

@article{TimdepPSSearch_BRAUN2010,
title = {{Time-dependent point source search methods in high energy neutrino astronomy}},
journal = {Astropart. Phys.},
volume = {33},
number = {3},
pages = {175-181},
year = {2010},
issn = {0927-6505},
doi = {https://doi.org/10.1016/j.astropartphys.2010.01.005},
url = {https://www.sciencedirect.com/science/article/pii/S0927650510000241},
author = {Jim Braun and others},
}

@article{annurev_Meszaros2017,
   author = "Mészáros, P.",
   title = {{Astrophysical sources of high-energy neutrinos in the IceCube era}}, 
   journal= "Annu. Rev. Nucl. Part. Sci.", 
   year = "2017",
   volume = "67",
   number = "Volume 67, 2017",
   pages = "45-67",
   doi = "https://doi.org/10.1146/annurev-nucl-101916-123304",
   url = "https://www.annualreviews.org/content/journals/10.1146/annurev-nucl-101916-123304",
   publisher = "Annual Reviews",
   issn = "1545-4134",
   type = "Journal Article",
}

@ARTICLE{GRBNu1997PRL,
       author = {{Waxman}, Eli and {Bahcall}, John},
        title = {{High energy neutrinos from cosmological gamma-ray burst fireballs}},
      journal = {\prl},
         year = 1997,
        month = mar,
       volume = {78},
       number = {12},
        pages = {2292-2295},
          doi = {10.1103/PhysRevLett.78.2292},
}

@ARTICLE{CCSNNu_Murase2011PRD,
       author = {{Murase}, Kohta and others},
        title = {{New class of high-energy transients from crashes of supernova ejecta with massive circumstellar material shells}},
      journal = {\prd},
         year = 2011,
        month = aug,
       volume = {84},
       number = {4},
          eid = {043003},
        pages = {043003},
          doi = {10.1103/PhysRevD.84.043003},
}

@article{Nakama2025GRB,
  title = {{Relaxation of time-variable neutron-loaded relativistic jets across the photosphere and their GeV-TeV neutrino counterparts}},
  author = {Nakama, Kanako and Kashiyama, Kazumi and Shimizu, Nobuhiro},
  journal = {Phys. Rev. D},
  volume = {113},
  issue = {10},
  pages = {103036},
  numpages = {13},
  year = {2026},
  month = {May},
  publisher = {American Physical Society},
  doi = {10.1103/4ryg-bzc7},
  url = {https://link.aps.org/doi/10.1103/4ryg-bzc7}
}

@article{Saurenhaus2026,
  title = {{Constraining the contribution of Seyfert galaxies to the diffuse neutrino flux in light of point source observations}},
  author = {Saurenhaus, Lena and others},
  journal = {Phys. Rev. D},
  volume = {113},
  issue = {2},
  pages = {023019},
  numpages = {21},
  year = {2026},
  month = {Jan},
  publisher = {American Physical Society},
  doi = {10.1103/f66p-k6z9},
  url = {https://link.aps.org/doi/10.1103/f66p-k6z9}
}

@article{Carpio_Seyfert_2026,
doi = {10.3847/1538-4357/ae7071},
url = {https://doi.org/10.3847/1538-4357/ae7071},
year = {2026},
month = {jun},
publisher = {The American Astronomical Society},
volume = {1005},
number = {1},
pages = {61},
author = {Carpio, Jose Alonso and Kheirandish, Ali and Murase, Kohta},
title = {{Multimessenger characterization of high-energy neutrino emission from the brightest neutrino-active galactic nuclei}},
journal = {ApJ}
}

@unknown{Eichmann2026BayesianSeyfert,
author = {Eichmann, Björn and others},
year = {2026},
month = {02},
pages = {},
title = {{Bayesian parameter study of the Seyfert-starburst composite galaxies NGC 1068 and NGC 7469}},
doi = {10.48550/arXiv.2602.15644}
}

@article{IceCube:ESTES2026,
doi = {10.3847/1538-4357/ae2c86},
url = {https://doi.org/10.3847/1538-4357/ae2c86},
year = {2026},
month = {feb},
publisher = {The American Astronomical Society},
volume = {998},
number = {1},
pages = {37},
author = {{IceCube Collaboration}},
title = {{Time-integrated southern-sky neutrino source searches with 10 yr of IceCube starting-track events at energies down to 1 TeV}},
journal = {ApJ}
}

@article{Wilks1938,
author = {S. S. Wilks},
title = {{The large-sample distribution of the likelihood ratio for testing composite hypotheses}},
volume = {9},
journal = {The Annals of Mathematical Statistics},
number = {1},
publisher = {Institute of Mathematical Statistics},
pages = {60 -- 62},
year = {1938},
doi = {10.1214/aoms/1177732360},
URL = {https://doi.org/10.1214/aoms/1177732360}
}

\end{document}